\documentclass[aps,prd,onecolumn,showpacs,showkeys,amsmath,amssymb]{revtex4}
\usepackage{graphicx}
\usepackage{dcolumn}
\usepackage{bm}
\begin{document}
\title{Chiral Properties of Baryon Fields with Flavor $SU(3)$ Symmetry}
\author{Hua-Xing Chen$^{1,2}$}
\email{hxchen@rcnp.osaka-u.ac.jp}
\author{V. Dmitra\v sinovi\' c$^3$}
\email{dmitrasin@yahoo.com}
\author{Atsushi Hosaka$^{1}$}
\email{hosaka@rcnp.osaka-u.ac.jp}
\author{Keitaro Nagata$^4$}
\email{nagata@phys.cycu.edu.tw}
\author{Shi-Lin Zhu$^{2}$}
\email{zhusl@phy.pku.edu.cn}
\affiliation{$^1$Research Center for Nuclear Physics, Osaka
University, Ibaraki 567--0047, Japan \\$^2$Department of Physics,
Peking University, Beijing 100871, China \\ $^3$ Vin\v ca Institute
of Nuclear Sciences, lab 010, P.O.Box 522, 11001 Beograd, Serbia \\
$^4$ Department of Physics, Chung-Yuan Christian University,
Chung-Li 320, Taiwan}
\begin{abstract}
We investigate chiral properties of local (non-derivative) fields
of baryons consisting of three quarks with flavor $SU(3)$
symmetry. We construct explicitly independent local three-quark
fields belonging to definite Lorentz and flavor representations.
Chiral symmetry is spontaneously broken and therefore the baryon
fields can have different chiral representations. It turns out
that the allowed chiral representations are strongly correlated
with the Lorentz group representations due to the color and
spatial structure of the local three-quark fields. We discuss some
implications of the allowed chiral symmetry representations on
phenomenological Lagrangians with chiral $U(3)_{\rm L} \otimes
U(3)_{\rm R}$ symmetry.
\end{abstract}
\maketitle
\pagenumbering{arabic}
%
\section{Introduction}
\label{sec:intro}
%

As the chiral symmetry of QCD is spontaneously broken, $SU(N_f)_L
\otimes SU(N_f)_R \rightarrow SU(N_f)_V$ ($N_f$ being the number of
flavors), the observed hadrons are classified by the residual
symmetry group representations of $SU(N_f)_V$. The full chiral
symmetry may then conveniently be represented by its non-linear
realization and this broken symmetry plays a dynamical role in the
presence of the Nambu-Goldstone bosons and their interactions.

Yet, as pointed out by Weinberg~\cite{Weinberg:1969hw}, there are
situations when it makes sense to consider algebraic aspects of
chiral symmetry, i.e. the chiral multiplets of hadrons. Such hadrons
may be classified in linear representations of the chiral symmetry
group with some representations mixing. One such situation becomes
realistic in the symmetry restored phase which is expected at high
temperatures and/or densities~\cite{Hatsuda:1994pi}. If hadrons
belong to certain representations of the chiral symmetry group,
certain physical properties such as the axial coupling constants are
determined by this symmetry. Therefore, the question as to what
chiral representations, possibly with mixing, the hadrons belong to
is of fundamental interest~\cite{lee72,lee81,Jido:2001nt}.

Another point of relevance is that the chiral representation can be
used as a measure of the internal structure of hadrons. For
instance, for a $\bar q q$ spin-one mesons, the possible chiral
representations are $(\mathbf{8}, \mathbf{1})$ and $(\mathbf{3},
\mathbf{\bar 3})$ and their left-right conjugates for flavor octet
mesons. As a matter of fact, for the multiquark hadrons, the allowed
chiral representations can be more complicated/higher dimensional
with increasing number of quarks and antiquarks. Hence the study of
chiral representations may provide some hints to the structure of
hadrons, extending possibly beyond the minimal constituent
picture~\cite{Jido:1997yk,Jido:1999hd,Beane:2002td,Benmerrouche:1989uc,Haberzettl:1998rw,DeTar:1988kn}.

Motivated by this argument, we have recently performed a complete
classification of baryon fields constructed from three quarks in the
local form with two light flavors (the so-called $SU(2)$
sector)~\cite{Nagata:2007di}. Such baryon fields are used as
interpolators for the study of two-point correlation functions in
the QCD sum rule approach and in the lattice
QCD~\cite{Ioffe:1981kw,Chung:1981cc,Espriu:1983hu,Lee:2002jb,Leinweber:2004it,Zanotti:2003fx}.
Strictly speaking, however, the chiral structure of an interpolator
does not directly reflect that of the physical state when chiral
symmetry is spontaneously broken. But the minimal configuration of
three quarks provides at least a guide to the simplest expectations
for baryons. Any deviation from such a simple structure may be an
indication of higher Fock-space components, such as the multi-quark
ones~\cite{Cohen:1996sb}.

Another reason for such a study of chiral classifications is related
to the number of independent fields. In principle, the correlation
functions should contain all information about physical states when
computed exactly. Practically, however, one must rely on some
approximation, and it has been observed in previous studies, that
the results may depend significantly on the choice of the
interpolators, which are generally taken as linear combinations of
the independent ones~\cite{Espriu:1983hu,Jido:2001nt,Chen:2006hy}.

In this paper, we perform a complete classification of baryon fields
written as local products (without derivatives) of three quarks
according to chiral symmetry group $SU(3)_L \otimes SU(3)_R$. This
is an extension of our previous work for the case of flavor
$SU(2)$~\cite{Nagata:2007di}. Technically, the $SU(3)$ algebra
introduces more complications, which makes insight less at work.
Hence, here we attempt to explore a rather technical aspect which
enables one to perform systematic classification. We derive general
transformation rules for baryon fields for the classification, while
maximally utilizing the Fierz transformations in order to implement
the Pauli principle among the quarks. The objective of $SU(3)$
baryon fields provides a simple but suitable exercise how the method
works. It can be extended to systems with more complex hadron fields
containing more quarks~\cite{Chen:2006hy}.

As in the previous paper~\cite{Nagata:2007di}, we first establish
the classification under the flavor $SU(3)$ symmetry, and then
investigate the properties under chiral symmetry group. The method
is based essentially on the tensor method for $SU(3)$, while the
Fierz method for the Pauli principle associated with the structure
in the color, flavor, Lorentz (spin) and orbital spaces is utilized
when establishing the independent fields. It turns out that for
local three-quark fields, the Pauli principle puts a constraint on
the structure of the Lorentz and chiral representations. This leads
essentially to the same permutation symmetry structure as in the
case of flavor $SU(2)$.

This paper is organized as follows. In Section~\ref{sec:fields}, we
establish the independent local baryon interpolating fields, and
investigate their flavor $SU(3)$ symmetry properties. In
Section~\ref{sec:chiral_transform}, we investigate the properties of
the baryon fields under chiral symmetry transformations $SU(3)_L
\otimes SU(3)_R$. We find that both flavor and chiral symmetry
properties are related to the structure of the Lorentz group.
Eventually, in Section~\ref{sec:chiral_representation}, we find that
this can be explained by the Pauli principle for the left and right
handed quarks, which puts a constraint on permutation symmetry
properties of three-quarks. Some complicated formulae are shown in
appendices.

%
\section{Flavor Symmetries of Three-Quark Baryon Fields}
\label{sec:fields}
%

Local fields for baryons consisting of three quarks
can be generally written as
\begin{eqnarray}
B(x) \sim \epsilon_{abc} \left(q^{aT}_A (x) C \Gamma_1
q^b_B(x)\right) \Gamma_2 q^c_C(x)\, , \label{def:field}
\end{eqnarray}
where $a,b,c$ denote the colour and $A,B,C$ the flavour indices, $C
= i\gamma_2 \gamma_0$ is the charge-conjugation operator,
$q_{A}(x)=(u(x),\;d(x),\;s(x))$ is the flavour triplet quark field
at location $x$, and the superscript $T$ represents the transpose of
the Dirac indices only (the flavour and colour $SU(3)$ indices are
{\it not} transposed). The antisymmetric tensor in color space
$\epsilon_{abc}$, ensures the baryons' being color singlets. The
space-time coordinate $x$ does nothing with our studies, and we
shall omit it. The matrices $\Gamma_{1,2}$ are Dirac matrices which
describe the Lorentz structure. With a suitable choice of
$\Gamma_{1,2}$ and taking a combination of indices of $A, B$ and
$C$, the baryon operators are defined so that they form an
irreducible representation of the Lorentz and flavour groups, as we
shall show in this section.

We employ the tensor formalism for flavour $SU(3)$ {\it a la}
Okubo~\cite{Okubo:1961jc,OSN83,MMOkubo65,Hara65,SchU65} for the
quark field $q$, although the explicit expressions in terms of $up$,
$down$ and $strange$ quarks are usually employed in lattice QCD and
QCD sum rule studies. We shall see that the tensor formulation
simplifies the classification of baryons into flavour multiplets and
leads to a straightforward (but generally complicated) derivation of
Fierz identities and chiral transformations of baryon operators.
This is in contrast with the $N_f = 2$ case where we explicitly
included isospin into the $\Gamma_{1,2}$ matrices and thus produced
isospin invariant/covariant objects~\cite{Nagata:2007di}. The reason
for this switch is that in the $N_f = 3$ case the baryons form
octets and decuplets, rather than doublets and quartets, but the
octet and decuplet projection operators cannot be easily introduced
into this formalism.

%
\subsection{Flavour $SU(3)_f$  decomposition for baryons}
\label{subsec:su3}
%

For the sake of notational completeness, we start with some
definitions. The quarks of flavor $SU(3)$ form either the
contra-variant ($\mathbf{3}$) or the covariant ($\mathbf{\bar 3}$)
fundamental representations. They are distinguished by either
upper or lower index as
\begin{eqnarray}
q^A \in q &=& \left(\begin{array}{c} u \\ d \\
s \end{array}\right)\, ,
\\ \nonumber
q_{A} \in q^\dagger  &=& (u^{*},\;d^{*},\;s^{*}) \, .
\label{def_qAqA}
\end{eqnarray}
The two conjugate fundamental representations transform
under flavor $SU(3)$ transformations as
\begin{eqnarray}
q &\to& \exp(i { \vec \lambda \over 2} \vec a){q} \, , \\ \nonumber
q^\dagger &\to& {q^\dagger }\exp(- i {\vec \lambda \over 2} \vec a)
\, ,
\label{eq:flavortransform}%
\end{eqnarray}%
where $a_N$ ($N=1,\cdots,8$) are the octet of $SU(3)_F$ group
parameters and $\lambda^N$ are the eight Gell-Mann matrices. Since
the latter are Hermitian, we may replace the transposed matrices
with the complex conjugate ones. The set of eight ${\bar \lambda}^N
= - (\lambda^{N})^T = - (\lambda^{N})^*$ matrices form the
generators of the irreducible $\mathbf{\bar 3}$ representation.

Now for three quarks, we show flavor $SU(3)$ irreducible
decomposition $\mathbf{3} \otimes \mathbf{3} \otimes \mathbf{3} =
\mathbf{1} \oplus \mathbf{8} \oplus \mathbf{8} \oplus \mathbf{10}$
explicitly in terms of three quarks. It can be done by making
suitable permutation symmetry representations of three-quark
products $q_A q_B q_C$.
\begin{enumerate}
\item
The totally antisymmetric combination which forms the singlet,
\begin{eqnarray} \Psi_{[ABC]} = {\cal N} \left(q_{A}q_{B}q_{C} +
q_{B}q_{C}q_{A} + q_{C}q_{A}q_{B} - q_{B}q_{A}q_{C} -
q_{A}q_{C}q_{B} - q_{C}q_{B}q_{A} \right) \, .
\end{eqnarray}
The normalization constant here is ${\cal N} = 1/\sqrt{6}$. In the
quark model this corresponds to $\Lambda(1405)$. In order to
represent this totally antisymmetric combination, we can use the
totally antisymmetric tensor $\epsilon^{ABC}$. Then the flavor
singlet baryon field $\Lambda$ can be written as:
\begin{equation}
\Lambda \equiv \epsilon^{ABC} \epsilon_{abc} \left( q^{aT}_A C
\Gamma_1 q^b_B \right) \Gamma_2 q^c_C \, .
\end{equation}

\item
The totally symmetric combination which forms the decuplet,
\begin{eqnarray}
\Psi_{\{ABC\}} = {\cal N} \left(q_{A}q_{B}q_{C} +
q_{B}q_{C}q_{A} + q_{C}q_{A}q_{B} + q_{B}q_{A}q_{C} +
q_{A}q_{C}q_{B} + q_{C}q_{B}q_{A} \right) \, .
\end{eqnarray}
The normalization constant depends on the set of quarks for baryons:
for $q_A, q_B, q_C = u, d, s, {\cal N} = 1/\sqrt{6}$, while it is
1/6 for $q_A, q_B, q_C = uuu$. In order to represent this totally
symmetric flavor structure, we introduce the totally symmetric
tensor $S_P^{ABC}$ ($P=1,\cdots,10$). Then the flavor decuplet
baryon field $\Delta$ can be written as:
\begin{equation}
\Delta^P \equiv S_P^{ABC} \epsilon_{abc} \left( q^{aT}_A C \Gamma_1
q^b_B \right) \Gamma_2 q^c_C \, .
\end{equation}
The non-zero components of $S_P^{ABC}$ ($=1$) are summarized in
Table~\ref{tab:SP_ABC}. The rest of components are just zero, for
instance, $S_1^{112}=0$.
\begin{table}[hbt]
\begin{center}
\caption{Non-zero components of $S_P^{ABC} ($=1$)$}
\begin{tabular}{c | c c c c c c c c c c}
\hline \hline $P$&1&2&3&4&5&6&7&8&9&10
\\ \hline $ABC$ & 111 & 112 & 113 & 122 & 123 & 133 & 222 & 223 & 233
& 333
\\ \hline Baryons & $\Delta^{++}$ & $\Delta^{+}$ & $\Sigma^{*+}$ &
$\Delta^{0}$ & $\Sigma^{*0}$ & $\Xi^{*0}$ & $\Delta^{-}$ &
$\Sigma^{*-}$ & $\Xi^{*-}$ & $\Omega^{-}$
\\ \hline
\end{tabular}
\label{tab:SP_ABC}
\end{center}
\end{table}

\item
The two mixed symmetry tensors of the $\rho$ and $\lambda$ types are
defined by
\begin{eqnarray} \label{def:octetrholambda}
\Psi_{[A\{B]C\}}^{\rho} &=& {\cal N} \left( 2q_{A}q_{B}q_{C} -
q_{B}q_{C}q_{A} - q_{C}q_{A}q_{B} - 2 q_{B}q_{A}q_{C} +
q_{A}q_{C}q_{B} + q_{C}q_{B}q_{A} \right) \, ,
\\ \nonumber
\Psi_{\{A[B\}C]}^{\lambda} &=& {\cal N} \left( 2 q_{A}q_{B}q_{C} -
q_{B}q_{C}q_{A} - q_{C}q_{A}q_{B} + 2 q_{B}q_{A}q_{C} -
q_{A}q_{C}q_{B} - q_{C}q_{B}q_{A} \right) \, .
\end{eqnarray}
Here the two symbols in $\{~\}$ are first symmetrized and then the
symbols in $[~]$ are anti-symmetrized. The normalization constant
depends again on the number of different kinds of terms. The
correspondence of the octet fields of (\ref{def:octetrholambda}) and
the physical ones can be made first by taking the following
combinations
\begin{eqnarray} \label{eq:rholambda}
N_{8\rho}^N &=& \epsilon^{ABD} (\bm{\lambda}^N)_{DC}
\Psi^{\rho}_{[A\{B]C\}} \, ,
\\
N_{8\lambda}^N &=& \epsilon^{BCD} (\bm{\lambda}^N)_{DA}
\Psi^{\lambda}_{\{A[B\}C]} \nonumber \, .
\end{eqnarray}
This kind of ``double index" ($DC$ for $N^N_{8\rho}$ and $DA$ for
$N^N_{8\lambda}$) notation for the baryon flavour has been used by
Christos~\cite{Christos:1986us}. In our discussions, we shall use
the following form for the flavor octet baryon field
\begin{eqnarray}
N^N &\equiv& \epsilon^{ABD} (\bm{\lambda}^N)_{DC} \epsilon_{abc}
\left( q^{aT}_A C \Gamma_1 q^b_B \right) \Gamma_2 q^c_C \, .
\label{def:octet_field}
\end{eqnarray}
It is of the $\rho$ type. But after using Fierz transformations to
interchange the second and the third quarks, the transformed one
contains $\lambda$ type also, as we shall show in the following. The
octet of physical baryon fields are then determined by
\begin{eqnarray} \label{def:octet_state}
& & N^1 \pm i N^2 \sim  \Sigma^{\mp} \, , \; \; \; N^3 \sim \Sigma^0
\, , \; \; \; N^8 \sim \Lambda \, ,
\\
\nonumber & & N^4 \pm i N^5 \sim \Xi^-, p\, ,\; \; \; N^6 \pm  i N^7
\sim \Xi^0 , n \, ,
\end{eqnarray}
or put into the $3 \times 3$ baryon matrix
\begin{eqnarray}
\mathfrak B = \left ( \begin{array} {c c c} {\Sigma^0 \over
\sqrt{2}} + { \Lambda^8 \over \sqrt{6}} & \Sigma^+ & p
\\ \Sigma^- & - {\Sigma^0 \over \sqrt{2}} + { \Lambda^8
\over \sqrt{6}} & n
\\ \Xi^- & \Xi^0 & - {2\over\sqrt{6}} \Lambda^8
\end{array} \right )
\label{eq:B} \, .
\end{eqnarray}
\end{enumerate}

%
\subsection{Counting the (in)dependent fields}
\label{subsec:baryon0}
%

In this section we investigate independent baryon fields for each
Lorentz group representation which is formed by three quarks. The
irreducible decomposition of the Lorentz group is done as
\begin{eqnarray} \left( (\frac12, 0) \oplus (0,\frac12)
\right)^3 \sim \left( (\frac12, 0) \oplus (0,\frac12) \right) \oplus
\left((1, \frac12) \oplus (\frac12,1) \right) \oplus \left(
(\frac32, 0) \oplus (0,\frac32) \right)\, ,
\end{eqnarray}
where we have ignored the multiplicity on the right hand side. The
three representations are described by the Dirac spinor field, the
Rarita-Schwinger's vector spinor field and the antisymmetric tensor
spinor field, respectively. In order to establish independent fields
when combined with color, flavor, and Lorentz (spin) degrees of
freedom, we employ the method of Fierz transformations which are
essentially equivalent to the use of the Pauli principle for three
quarks. Here we demonstrate the essential idea for the simplest case
for the Dirac spinor, $ (\frac12, 0) \oplus (0,\frac12) $. Other
cases are briefly explained in Appendices~\ref{app:baryon1} and
\ref{app:baryon2}.

\subsubsection{Flavor singlet baryon}
Let us start with writing down five baryon fields which contain a diquark formed by
five sets of Dirac matrices,
$1, \gamma_5, \gamma_\mu, \gamma_\mu \gamma_5$ and $\sigma_{\mu \nu}$,
\begin{eqnarray}
\begin{array}{l} \Lambda_1 = \epsilon_{abc}
\epsilon^{ABC} (q_A^{aT} C q_B^b) \gamma_5 q_C^c \, ,\\
\Lambda_2 = \epsilon_{abc} \epsilon^{ABC} (q_A^{aT} C \gamma_5
q_B^b) q_C^c \, ,\\
\Lambda_3 = \epsilon_{abc} \epsilon^{ABC} (q_A^{aT} C \gamma_\mu
\gamma_5 q_B^b) \gamma^\mu q_C^c \, ,\\
\Lambda_4 = \epsilon_{abc} \epsilon^{ABC} (q_A^{aT} C \gamma_\mu
q_B^b) \gamma^\mu \gamma_5 q_C^c  \, ,\\
\Lambda_5 = \epsilon_{abc} \epsilon^{ABC} (q_A^{aT} C
\sigma_{\mu\nu} q_B^b) \sigma_{\mu\nu} \gamma_5 q_C^c  \, .\
\end{array}
\end{eqnarray}
Among these five fields, we can show that the fourth and fifth ones
vanish, $\Lambda_{4,5}=0$. This is due to the Pauli principle
between the first two quarks, and can be verified, for instance, by
taking the transpose of the diquark component and compare the
resulting three-quark field with the original
expressions~\cite{Christos:1986us}. The Pauli principle can also be
used between the first and the third quarks, so we construct the
primed fields where the second and the third quarks are
interchanged, for instance,
\begin{eqnarray}\nonumber
\Lambda^\prime_{1} = \epsilon_{abc} \epsilon^{ABC} (q_A^{aT} C
q_C^c) \gamma_5 q_B^b\, .
\end{eqnarray}
Now expressing $\Lambda_{i}$ in terms of the Fierz transformed
fields $\Lambda^\prime_{i}$, we find the following relations (see
Appendix~\ref{app:fierz}),
\begin{eqnarray}\nonumber
\begin{array}{l}
\Lambda_1 = - {1 \over 4} \Lambda^\prime_1 - {1 \over 4}
\Lambda^\prime_2 - {1 \over 4} \Lambda^\prime_3\, , \\
\Lambda_2 = - {1 \over 4} \Lambda^\prime_1 - {1 \over 4}
\Lambda^\prime_2 + {1 \over 4} \Lambda^\prime_3 \, , \\
\Lambda_3 = - \Lambda^\prime_1 + \Lambda^\prime_2 + {1 \over 2}
\Lambda_3^\prime \, .
\end{array}
\label{fiveLambda}
\end{eqnarray}
On the other hand, by changing the indices $B, C$ and $b, c$, for
instance,
\begin{eqnarray}\nonumber
\Lambda^\prime_{1} &=& \epsilon_{acb} \epsilon^{ACB} (q_A^{aT} C
q_B^b) \gamma_5 q_C^c \\ \nonumber &=& \epsilon_{abc} \epsilon^{ABC}
(q_A^{aT} C q_B^b) \gamma_5 q_C^c \, ,
\end{eqnarray}
we see that the primed fields are just the corresponding unprimed
ones, $ \Lambda^\prime_i = \Lambda_i$. Consequently, we obtain three
homogeneous linear equations whose rank is just one, and we find the
following solution
\begin{equation} \nonumber
\Lambda_3 = 4 \Lambda_2 = - 4 \Lambda_1 \, , \Lambda_4 = \Lambda_5 =
0 \, .
\end{equation}
We see that there is only one non-vanishing independent field, which
in the quark model corresponds to the odd-parity $\Lambda(1405)$.

\subsubsection{The flavour decuplet baryons}

Among the five decuplet baryon fields formed by the five different
$\gamma$-matrices, only two are non-zero:
\begin{eqnarray}
\begin{array}{l}
\Delta^P_4 = \epsilon_{abc} S_P^{ABC} (q_A^{aT} C \gamma_\mu
q_B^b) \gamma^\mu \gamma_5 q_C^c \, ,\\
\Delta^P_5 = \epsilon_{abc} S_P^{ABC} (q_A^{aT} C \sigma_{\mu\nu}
q_B^b) \sigma_{\mu\nu} \gamma_5 q_C^c \, .
\end{array}
\label{fiveDelta}
\end{eqnarray}
Performing the Fierz transformation and with the relation
$\Delta^{P\prime}_i = - \Delta^P_i$ ($\epsilon_{acb} S_P^{ACB} = -
\epsilon_{abc} S_P^{ABC}$), we find that there is only a trivial
(null) solution to the homogeneous linear equations. Therefore, the
Dirac baryon fields (fundamental representation of the Lorentz
group) formed by three quarks can not survive the flavor decuplet.

\subsubsection{The flavor octet baryon fields}
\label{subsub:octet}

Let us start once again with five fields, which have three
potentially non-zero ones
\begin{eqnarray}
\begin{array}{l} N_1^N = \epsilon_{abc}
\epsilon^{ABD} \lambda_{DC}^N (q_A^{aT} C q_B^b) \gamma_5 q_C^c \, ,\\
N_2^N = \epsilon_{abc} \epsilon^{ABD} \lambda_{DC}^N (q_A^{aT} C
\gamma_5 q_B^b) q_C^c \, ,\\
N_3^N = \epsilon_{abc} \epsilon^{ABD} \lambda_{DC}^N (q_A^{aT} C
\gamma_\mu \gamma_5 q_B^b) \gamma^\mu q_C^c \, ,\\
N_4^N = \epsilon_{abc} \epsilon^{ABD} \lambda_{DC}^N (q_A^{aT} C
\gamma_\mu q_B^b) \gamma^\mu \gamma_5 q_C^c = 0 \, ,\\
N_5^N = \epsilon_{abc} \epsilon^{ABD} \lambda_{DC}^N (q_A^{aT} C
\sigma_{\mu\nu} q_B^b) \sigma_{\mu\nu} \gamma_5 q_C^c = 0 \, .\\
\end{array}
\label{eq:fiveNs}
\end{eqnarray}
These octet baryon fields have been studied in
Refs~\cite{Ioffe:1981kw,Espriu:1983hu,Chung:1981cc}, where the
independent ones are clarified. As before, we perform the Fierz
rearrangement to obtain five equations with the primed fields, while
$N_4^{N\prime}$ and $N_5^{N\prime}$ are not zero. For the first
three equations, $N_{1,2,3}$ on the left hand side should be
expressed by the primed fields. To this end,  we can use the Jacobi
identity
\begin{eqnarray}
\epsilon^{ABD} \lambda_{DC}^N + \epsilon^{BCD} \lambda_{DA}^N +
\epsilon^{CAD} \lambda_{DB}^N = 0 \, , \label{eq:Jacobi}
\end{eqnarray}
which can be used to relate the original fields $N^N_i$ and primed
ones $N^{N\prime}_i$, for instance,
\begin{eqnarray}
\nonumber \left ( \epsilon^{ABD} \lambda_{DC}^N + \epsilon^{BCD}
\lambda_{DA}^N + \epsilon^{CAD} \lambda_{DB}^N \right ) (q_A^{aT} C
q_B^b) \gamma_5 q_C^c = 0 \, ,
\end{eqnarray}
 from which we find
 \begin{eqnarray} \nonumber
N_1^{N\prime} = - {1 \over 2} N_1^{N} \, ,
\end{eqnarray}
and the same relations for $N_{2,3}^N$. There are no relations
between $N_{4,5}^{N}$ and $N_{4,5}^{N\prime}$. Altogether, we have
five equations. The equations related to $N_4^{N}$ and $N_5^{N}$ are
also necessary because the corresponding primed ones are not zero.
They can be solved to obtain the following solutions:
\begin{equation} \nonumber
{2 \over 3} N_4^{N\prime} = N^N_3 = N^N_1 - N^N_2 \, , N_5^{N\prime}
= - 3 (N^N_1 + N^N_2) \, ,
\end{equation}
which indicates that there are two independent octet fields, for
instance, $N^N_1$ and $N^N_2$. Thus we have shown the same result
just as in the two-flavour case~\cite{Nagata:2007di}. In the
following sections we shall show that the difference between the two
fields $N_1$ and $N_2$ lies in their chiral properties: $N^N_1 -
N^N_2$ together with $\Lambda$ belong to $(\mathbf{\bar
3},~\mathbf{3}) \oplus (\mathbf{3},~\mathbf{\bar 3})$, and the other
$N^N_1 + N^N_2$ belongs to $(\mathbf{8},~\mathbf{1}) \oplus
(\mathbf{1},~\mathbf{8})$.

There are two ways to construct the octet baryon fields. One is done
already as shown in Eqs.~(\ref{eq:fiveNs}), whose flavor structure
is the same as the $\rho$ type baryon field $N^N_{8\rho}$ in
Eqs.~(\ref{eq:rholambda}):
\begin{eqnarray}
\mathbf{3}\otimes\mathbf{3}\otimes\mathbf{3} \longrightarrow
(\mathbf{3}\otimes\mathbf{3})\otimes\mathbf{3} \longrightarrow
\mathbf{\bar 3} \otimes \mathbf{3} \longrightarrow \mathbf{8}_\rho
\, .\label{def:8rho}
\end{eqnarray}
The other $\lambda$ type baryon field $N^N_{8\lambda}$ is
complicated when used straightforwardly:
\begin{eqnarray}
\mathbf{3}\otimes\mathbf{3}\otimes\mathbf{3} \longrightarrow
(\mathbf{3}\otimes\mathbf{3})\otimes\mathbf{3} \longrightarrow
\mathbf{6} \otimes \mathbf{3} \longrightarrow \mathbf{8}_\lambda \,
.\label{def:8lambda}
\end{eqnarray}
Therefore, we use another way based on
\begin{eqnarray}
\mathbf{3}\otimes\mathbf{3}\otimes\mathbf{3} \longrightarrow
\mathbf{3}\otimes(\mathbf{3}\otimes\mathbf{3}) \longrightarrow
\mathbf{3} \otimes \mathbf{\bar 3} \longrightarrow
\mathbf{8}^\prime_\rho \, .\label{def:8rhoprime}
\end{eqnarray}
This contains partly $\mathbf{8}_\lambda$, and it is easily to
verify that (\ref{def:8rho}) and (\ref{def:8rhoprime}) compose a
full description of octet baryon which is also fully described by
using (\ref{def:8rho}) and (\ref{def:8lambda}). The way
$\mathbf{8}_\rho$ leads to octet fields $N^N_i$, and the other way
$\mathbf{8}^\prime_\rho$ leads to other five ones
\begin{eqnarray}
\begin{array}{l}
\widetilde{N}_1^{N} = \epsilon_{abc} \epsilon^{ACD} \lambda_{DB}^N
(q_A^{aT} C q_B^b) \gamma_5 q_C^c \, ,
\\
\widetilde{N}_2^{N} = \epsilon_{abc} \epsilon^{ACD} \lambda_{DB}^N
(q_A^{aT} C \gamma_5 q_B^b) q_C^c \, ,
\\ \widetilde{N}_3^{N} = \epsilon_{abc} \epsilon^{ACD} \lambda_{DB}^N
(q_A^{aT} C \gamma_\mu \gamma_5 q_B^b) \gamma^\mu q_C^c \, ,
\\
\widetilde{N}_4^{N} = \epsilon_{abc} \epsilon^{ACD} \lambda_{DB}^N
(q_A^{aT} C \gamma_\mu q_B^b) \gamma^\mu \gamma_5 q_C^c \, ,
\\
\widetilde{N}_5^{N} = \epsilon_{abc} \epsilon^{ACD} \lambda_{DB}^N
(q_A^{aT} C \sigma_{\mu\nu} q_B^b) \sigma_{\mu\nu} \gamma_5 q_C^c \,
.
\end{array}
\end{eqnarray}
However, these fields can be related to the previous ones by
changing the flavor and color indices $B,C$ and $b,c$:
\begin{equation} \nonumber
\widetilde{N}_i^{N} = - N_{i}^{N\prime} \, . \label{eq:secondoctet}
\end{equation}
In nearly all the cases, the octet baryon fields from the second way
can be related to the ones from the first way. Therefore, we shall
omit the discussion of the second octet. One exception which
concerns the chiral representation $(\mathbf{\bar
3},~\mathbf{3})\otimes(\mathbf{6},~\mathbf{3})$ is discussed in
Appendix~\ref{app:chiral_rep}.

%
\subsection{A short summary of independent baryon fields}
\label{subsec:sum_baryon}
%
Properties of baryons fields expressed by the Rarita-Schwinger
fields with one Lorentz index and those of the antisymmetric
tensor-spinor fields with two Lorentz indices are discussed in
Appendices~\ref{app:baryon1} and \ref{app:baryon2}, respectively.
Here let us make a short summary for independent baryon fields for
all cases constructed by three quarks. For simplicity, here we
suppress the antisymmetric tensor in color space $\epsilon_{abc}$,
since it appears in all baryon fields in the same manner.
Furthermore, it is convenient to introduce a ``tilde-transposed"
quark field $\widetilde{q}$ as follows
\begin{eqnarray}
\widetilde{q}=q^T C \gamma_5 \, . \label{def:tilde}
\end{eqnarray}

As we have shown already, for Dirac fields without Lorentz index,
there are one singlet field $\Lambda$ and two octet fields $N^N_1$
and $N^N_2$:
\begin{eqnarray} \nonumber
\Lambda_1 &=& \epsilon^{ABC} (\widetilde{q}_A \gamma_5 q_B) \gamma_5
q_C \label{def:lambda} \, ,\\ \nonumber N^N_1 &=& \epsilon^{ABD}
\lambda_{DC}^N (\widetilde{q}_A \gamma_5 q_B) \gamma_5 q_C
\label{def:n1} \, ,
\\ \nonumber N^N_2 &=& \epsilon^{ABD} \lambda_{DC}^N (\widetilde{q}_A
q_B) q_C \label{def:n2} \, .
\end{eqnarray}

For the Rarita-Schwinger fields with one Lorentz index, we would
consider one singlet, three octet and one decuplet fields:
\begin{eqnarray}
\nonumber \Lambda_{1\mu} &=& \epsilon^{ABC} (\widetilde{q}_A
\gamma_5 q_B) \gamma_\mu q_C \, ,\\ \nonumber N^{N}_{1\mu} &=&
\epsilon^{ABD}
\lambda_{DC}^N (\widetilde{q}_A \gamma_5 q_B) \gamma_\mu q_C \, , \\
\nonumber N^{N}_{2\mu} &=& \epsilon^{ABD} \lambda_{DC}^N
(\widetilde{q}_A q_B) \gamma_\mu \gamma_5 q_C \, , \\
\nonumber N^{N}_{3\mu} &=& - \epsilon^{ABD} \lambda_{DC}^N
(\widetilde{q}_A \gamma_\mu q_B) \gamma_5 q_C \, ,
\\ \nonumber \Delta^P_{5\mu} &=& - S_P^{ABC} (\widetilde{q}_A \gamma_\mu
\gamma_5 q_B) q_C \, .
\end{eqnarray}
However, we find that $\Lambda_{1\mu} = \gamma_\mu \gamma_5
\Lambda$, $N^N_{1\mu} = \gamma_\mu \gamma_5 N^N_1$ and $N^N_{2\mu} =
\gamma_\mu \gamma_5 N^N_2$. So, there are two non-vanishing
independent fields: one octet field $N^N_\mu$ and one decuplet field
$\Delta_\mu$. By using the projection operator:
\begin{eqnarray}
P^{3/2}_{\mu\nu} = (g_{\mu\nu} - {1\over4}\gamma_\mu\gamma_\nu) \, ,
\end{eqnarray}
they can be written as
\begin{eqnarray} \nonumber
N^N_\mu = P^{3/2}_{\mu\nu} N^{N}_{3\nu} &=& - (g_{\mu\nu} -
{1\over4}\gamma_\mu\gamma_\nu) \epsilon^{ABD} \lambda_{DC}^N
(\widetilde{q}_A \gamma_\mu q_B) \gamma_5 q_C
\\ \nonumber &=& N^N_{3\mu} + {1\over4} \gamma_\mu \gamma_5 (N^N_1 -
N^N_2) \, , \label{def:n3}
\\ \nonumber \Delta^P_{\mu} = P^{3/2}_{\mu\nu} \Delta^P_{5\nu} &=& - (g_{\mu\nu} -
{1\over4}\gamma_\mu\gamma_\nu) S_P^{ABC} (\widetilde{q}_A \gamma_\mu
\gamma_5 q_B) q_C \\ \nonumber &=& \Delta^P_{5\mu} \, .
\label{def:delta1}
\end{eqnarray}

For tensor fields with two antisymmetric Lorentz indices, we would
have one singlet, three octet and two decuplet fields:
\begin{eqnarray} \nonumber
\Lambda_{1\mu} &=& \epsilon^{ABC} (\widetilde{q}_A \gamma_5 q_B)
\sigma_{\mu\nu} \gamma_5 q_C \, ,\\ \nonumber N^N_{3\mu\nu} &=& -
\epsilon^{ABD} \lambda_{DC}^N (\widetilde{q}_A \gamma_{\mu} q_B)
\gamma_\nu q_C + ( \mu \leftrightarrow \nu ) \, ,
\\ \nonumber N^N_{10\mu\nu} &=& \epsilon^{ABD} \lambda_{DC}^N
(\widetilde{q}_A \gamma_5 q_B) \sigma_{\mu\nu} \gamma_5 q_C \, ,\\
\nonumber N^N_{11\mu\nu} &=& \epsilon^{ABD} \lambda_{DC}^N
(\widetilde{q}_A q_B) \sigma_{\mu\nu} q_C \, ,
\\ \nonumber \Delta^P_{2\mu\nu} &=& - S_P^{ABC} (\widetilde{q}_A \gamma_{\mu}
\gamma_5 q_B) \gamma_\nu \gamma_5 q_C + ( \mu \leftrightarrow \nu )
\, ,
\\ \nonumber \Delta^P_{7\mu\nu} &=& S_P^{ABC} (\widetilde{q}_A \sigma_{\mu\nu} \gamma_5 q_B) \gamma_5
q_C \, .
\end{eqnarray}
But in this case, we can show that there is only one non-vanishing
field $\Delta_{\mu\nu}$:
\begin{eqnarray}
\nonumber \Delta^P_{\mu\nu} = \Gamma^{\mu\nu\alpha\beta}
\Delta^P_{7\mu\nu} &=& \Gamma^{\mu\nu\alpha\beta} S_P^{ABC}
(\widetilde{q}_A \sigma_{\mu\nu} \gamma_5 q_B) \gamma_5 q_C
\\ \nonumber &=& \Delta^P_{7\mu\nu} - {i\over2} \gamma_\mu \gamma_5
\Delta^P_{5\nu} + {i\over2} \gamma_\nu \gamma_5 \Delta^P_{5\mu} \, ,
\label{def:delta2}
\end{eqnarray}
where
\begin{eqnarray}
\Gamma^{\mu\nu\alpha\beta} = ( g^{\mu\alpha}g^{\nu\beta} -
{1\over2}g^{\nu\beta}\gamma^{\mu}\gamma^\alpha +
{1\over2}g^{\mu\beta}\gamma^\nu\gamma^\alpha +
{1\over6}\sigma^{\mu\nu}\sigma^{\alpha\beta})\, .
\end{eqnarray}

%
\section{Chiral Transformations}
\label{sec:chiral_transform}
%

In this section, we establish the chiral transformation properties
of the baryon fields which we have obtained in the previous section.
Technically, this requires somewhat complicated algebra. However,
the final result will be understood by making the left and right
handed decomposition, as we will perform in the next section.

Let us start with the chiral transformation properties of quarks
which are given by the following equations:
\begin{eqnarray}\nonumber
\bf{U(1)_{V}} &:&  q \to \exp(i {\lambda^0 \over 2} a_{0}) q  = q +
\delta q
\label{eq:u1v} \, , \\
\bf{SU(3)_V} &:& {q} \to \exp (i {\vec \lambda \over 2} \vec a ){q}
=
q + \delta^{\vec{a}} q \label{eq:su3v} \, , \\
\nonumber\bf{U(1)_{A}} &:& q \to \exp(i \gamma_5 {\lambda^0 \over 2}
b_{0}) q = q + \delta_5 q \label{e:u1a} \, , \\ \nonumber
\bf{SU(3)_A} &:& {q} \to \exp (i \gamma_{5} {\vec \lambda \over 2}
\vec b){q} = q + \delta_5^{\vec{b}} q \label{eq:su3a} \, ,
\end{eqnarray}
where $\lambda^0 = \sqrt{2/3} \mathbf{1}$, $a^0$ is an infinitesimal
parameter for the $U(1)_V$ transformation, $\vec{a}$ the octet of
$SU(3)_V$ group parameters, $b^0$ an infinitesimal parameter for the
$U(1)_A$ transformation, and $\vec{b}$ the octet of the chiral
transformations.

The $U(1)_V$ chiral transformation is trivial which picks up a phase
factor proportional to the baryon number. The $U(1)_A$ chiral
transformation is slightly less trivial, and the baryon fields are
transformed as
\begin{eqnarray} \nonumber\label{eq:U1Atransform}
\delta_5 \Lambda &=& - i \gamma_5 \sqrt{1\over6} b^0 \Lambda \, ,
\\ \nonumber \delta_5 (N^N_1 - N^N_2) &=& - i \gamma_5 \sqrt{1\over6} b^0 ( N^N_1 - N^N_2 ) \, ,
\\ \nonumber \delta_5 (N^N_1 + N^N_2) &=& i \gamma_5 \sqrt{3\over2} b^0 ( N^N_1 + N^N_2
) \, ,
\\ \delta_5 N^N_\mu &=& i \gamma_5 \sqrt{1\over6} b^0 N^N_{\mu} \, ,
\\ \nonumber \delta_5 \Delta^P_\mu &=& i \gamma_5 \sqrt{1\over6} b^0 \Delta^P_{\mu} \, ,
\\ \nonumber \delta_5 \Delta^P_{\mu\nu} &=& i \gamma_5 \sqrt{3\over2} b^0
\Delta^P_{\mu\nu} \, .
\end{eqnarray}
We note that the combinations of $N_1^N \pm N_2^N$ form different
representations.

Under the vector chiral transformation, the baryon fields are
transformed as
\begin{eqnarray} \nonumber\label{eq:vectortransform}
\delta^{\vec{a}} \Lambda &=& 0 \, ,
\\ \nonumber \delta^{\vec{a}} N^N_1 &=& - a^M f^{NMO} N^O_1 \, ,
\\ \nonumber \delta^{\vec{a}} N^N_2 &=& - a^M f^{NMO} N^O_2 \, ,
\\ \delta^{\vec{a}} N^N_\mu &=& - a^M f^{NMO} N^N_\mu \, ,
\\ \nonumber \delta^{\vec{a}} \Delta^P_\mu &=& { 3 i \over 2} a^M g_7^{PMQ} \Delta^Q_{\mu} \, ,
\\ \nonumber \delta^{\vec{a}} \Delta^P_{\mu\nu} &=& {3 i \over2} a^M g_7^{PMQ} \Delta^Q_{\mu\nu}
\, ,
\end{eqnarray}
where $f^{ABC}$ is the standard antisymmetric structure constant of
$SU(3)$, and $g^{ABC}_7$ is defined in Table~\ref{tab:couplings}.
Eqs.~(\ref{eq:vectortransform}) show nothing but the flavor charge
of the baryons. For example, we can show explicitly:
\begin{eqnarray}\nonumber
\delta^{a3} p = + {i\over2} a_3 p \, , \delta^{a3} n = - {i\over2}
a_3 n \, , \delta^{a3} \Delta^{++} = {3 i\over2} a_3 \Delta^{++} \,
\cdots
\end{eqnarray}

The transformation rule under the axial-vector chiral
transformations are rather complicated as they are no longer
conserved and reflect the internal structure of baryons. To start
with, we have the axial transformation of the three-quark baryon
fields such as
\begin{eqnarray} \nonumber
\delta_5^{\vec b} \Lambda &=& \epsilon_{abc} \epsilon^{ABC} \Big (
(q_A^{aT} C q_B^b) \gamma_5 ( \delta_5^{\vec b} q_C^c ) + (q_A^{aT}
C (\delta_5^{\vec b} q_B^b)) \gamma_5  q_C^c + ( (\delta_5^{\vec b}
q_A^{aT}) C q_B^b) \gamma_5 q_C^c \Big ) \, .
\end{eqnarray}
The calculation is complicated, but rather straightforward. Here, we
show therefore the final result of the axial transformation:
\begin{eqnarray} \nonumber\label{eq:axialtransform}
\delta_5^{\vec{b}} \Lambda &=& {i\over2} \gamma_5 b^N (N^N_1 -
N^N_2) \, ,
\\ \nonumber \delta_5^{\vec{b}} (N^N_1 - N^N_2) &=& {4 i \over 3} \gamma_5 b^N \Lambda +
i \gamma_5 b^M d^{NMO} ( N^O_1 - N^O_2 ) \, ,
\\ \nonumber \delta_5^{\vec{b}} (N^N_1 + N^N_2) &=& - \gamma_5 b^M f^{NMO} ( N^O_1 + N^O_2
) \, ,
\\ \delta_5^{\vec{b}} N^N_\mu &=& i \gamma_5 b^M ( d^{MNO} - {2i \over 3} f^{MNO}
) N^O_{\mu} + i \gamma_5 b^M g_3^{MNP} \Delta^P_{\mu} \, ,
\\ \nonumber \delta_5^{\vec{b}} \Delta^P_\mu &=& - 2 i \gamma_5 b^M
g_5^{PMO} N^O_{\mu} + {i\over2} \gamma_5 b^M g_7^{PMQ}
\Delta^Q_{\mu} \, ,
\\ \nonumber \delta_5^{\vec{b}} \Delta^P_{\mu\nu} &=& {3 i\over2} \gamma_5 b^M g_7^{PMQ}
\Delta^Q_{\mu\nu} \, .
\end{eqnarray}
The coefficients $d^{ABC}$ are the standard symmetric structure
constants of $SU(3)$. For completeness, we show the following
equation which define the $d$ and $f$ coefficients
\begin{eqnarray}
\lambda^N_{AB} \lambda^M_{BC} &=& (\lambda^N \lambda^M)_{AC} =
{1\over2} \{ \lambda^N, \lambda^M \}_{AC} + {1\over2} [ \lambda^N,
\lambda^M ]_{AC}
\\ \nonumber &=& {2\over3} \delta^{NM} \delta_{AC} + (d^{NMO}
+ i f^{NMO}) \lambda^O_{AC} \, .
\end{eqnarray}
Furthermore, the following formulae define the coefficients $g_3$,
$g_5$ and $g_7$, which are proved by using $Mathematica$, a software
good at matrix calculation:
\begin{eqnarray} \label{eq:coupling}
&& \epsilon^{ADE} \lambda^N_{DB} \lambda^M_{EC} = g_1^{NMO}
\epsilon^{ABD} \lambda^O_{DC} + g_2^{NMO} \epsilon^{ACD}
\lambda^O_{DB} + g_3^{NMP} S_P^{ABC} + g_4^{NM} \epsilon^{ABC} \, ,
\\ \nonumber && S_Q^{ABD} \lambda^M_{DC} = g_5^{QMO}
\epsilon^{ABD} \lambda^O_{DC} + g_6^{QMO} \epsilon^{ACD}
\lambda^O_{DB} + g_7^{QMP} S_P^{ABC} + g_8^{QM} \epsilon^{ABC} \, ,
\end{eqnarray}
where indices $A \sim E$ take values 1, 2 and 3, $N$, $M$ and $O$
$1,\cdots,8$, and $P$ and $Q$  $1,\cdots,10$. The coefficients
$g_3$, $g_5$ and $g_7$ are listed in Table~\ref{tab:couplings},
where we use ``0'' instead of ``10''. Other coefficients can be
related to $d$, $f$, $g_3$, $g_5$ and $g_7$:
\begin{eqnarray} \nonumber\label{def:coefficient}
g_1^{MNO} &=& - d^{MNO} - {i\over3}f^{MNO} \, ,
\\ \nonumber g_2^{MNO} &=& d^{MNO} - {i\over3}f^{MNO} \, ,
\\ g_4^{MN} &=& - {1\over3} \delta^{MN} \, ,
\\ \nonumber g_6^{QMO} &=& - 2 g_5^{QMO} \, ,
\\ \nonumber g_8^{MN} &=& 0 \, .
\end{eqnarray}
Let us explain Eqs.~(\ref{eq:coupling}) a bit more. The quantities
on the left hand side have three indices $A$, $B$ and $C$, and
therefore, they are regarded as direct products of three fundamental
representations of $SU(3)$:
$\mathbf{3}\otimes\mathbf{3}\otimes\mathbf{3}$. They can be
decomposed into irreducible components by applying the four kinds of
operators: $\epsilon_{ABC}$, $\epsilon^{ABD} \lambda^O_{DC}$,
$\epsilon^{ACD} \lambda^O_{DB}$ and $S_P^{ABC}$, which correspond to
$\mathbf{1}$, $\mathbf{8}$, $\mathbf{8}$ and $\mathbf{10}$ of
$SU(3)$, respectively.

Eqs.~(\ref{eq:U1Atransform}) and (\ref{eq:axialtransform}) imply
that $\Lambda$ and $N^N_1 - N^N_2$ are together combined into one
chiral multiplet, and $N^N_\mu$ and $\Delta^P_\mu$ are together
combined into another chiral multiplet. While $N^N_1 + N^N_2$ and
$\Delta^P_{\mu\nu}$ are transformed into themselves under chiral
transformation. In our following discussion, we will find that
$\Lambda$ and $N^N_1 - N^N_2$ belong to the chiral representation
$(\mathbf{\bar 3}, \mathbf{3}) \oplus (\mathbf{3}, \mathbf{\bar
3})$, $N^N_1 + N^N_2$ belongs to the chiral representation
$(\mathbf{8}, \mathbf{1}) \oplus (\mathbf{1}, \mathbf{8})$,
$N^N_\mu$ and $\Delta^P_\mu$ belong to the chiral representation
$(\mathbf{6}, \mathbf{3}) \oplus (\mathbf{3}, \mathbf{6})$, and
$\Delta^P_{\mu\nu}$ belongs to the chiral representation
$(\mathbf{10}, \mathbf{1}) \oplus (\mathbf{1}, \mathbf{10})$. We
show several examples of the axial-vector chiral transformation:
\begin{eqnarray}
\nonumber \delta_5^{b3} p_- = {i\over2}\gamma_5 b_3 p_- \, ,
\delta_5^{b3} p_+ = {i\over2} \gamma_5 b_3 p_+ \, , \delta_5^{b3}
p_\mu = {5i\over6} \gamma_5 b_3 p_\mu - {4i\over3} \gamma_5 b_3
\Delta^+_\mu \, ,
\end{eqnarray}
where $p_-$ belongs to the octet baryon fields $N_1^N - N_2^N$,
$p_+$ belongs to $N^N_1 + N^N_2$, and $p_\mu$ belongs to $N^N_\mu$
(see Eqs.~(\ref{def:octet_state})).

\begin{table}[tbh]
\begin{center}
\caption{$g$-coefficients defined by Eqs.~(\ref{eq:coupling})}
\begin{tabular}{|c|c|c|c|c|c|c|c|c|}
\hline $g_3$ & \multicolumn{7}{c|}{133, 138, 144, 146, 254, 256,
272, 279, 439, 463, 468, 573, 578, 612, 619, 636} & $-1/3$
\\ \cline{2-9} & \multicolumn{7}{c|}{162, 169, 313, 318, 349, 366,
414, 416, 524, 526, 643, 648, 722, 729, 753, 758} & $1/3$
\\ \cline{2-9} & \multicolumn{7}{c|}{154, 179, 215, 233, 246, 269,
328, 359, 376, 424, 455, 478, 516, 563, 622, 658, 712, 743, 765} &
$-i/3$
\\ \cline{2-9} & \multicolumn{7}{c|}{125, 156, 172, 238, 244, 262, 323,
426, 473, 514, 539, 545, 568, 629, 653, 675, 719, 736, 748} & $i/3$
\\ \cline{2-9} & 183, 686, 818, 835, 849 & $-1/\sqrt{3}$
& \multicolumn{3}{c|}{167, 251, 277, 411, 570, 640} & $-1$ & 342,
364 & $-2/3$
\\ \cline{2-9} & 188, 385, 489, 813, 866 & $1/\sqrt{3}$
& \multicolumn{3}{c|}{141, 460, 521, 617, 727, 750} & $1$ & 432, 634
& $2/3$
\\ \cline{2-9} & 283, 288, 589, 876 & $- i/\sqrt{3}$
& \multicolumn{3}{c|}{177, 421, 470, 511, 560, 627} & $-i$ & 352,
374 & $-2i/3$
\\ \cline{2-9} & 786, 823, 828, 859 & $i/\sqrt{3}$
& \multicolumn{3}{c|}{151, 241, 267, 650, 717, 740} & $i$ & 532, 734
& $2i/3$
\\ \hline \hline $g_5$ & \multicolumn{3}{c|}{125, 141, 227, 261, 313, 346,
357, 414, 425, 614, 625} & $1/6$ & \multicolumn{3}{c|}{318, 668,
881, 984} & $1/2\sqrt{3}$
\\ \cline{6-9} & \multicolumn{3}{c|}{663, 716, 727, 813, 846, 857, 927, 943, 961, 057, 064}
& & \multicolumn{3}{c|}{381, 686, 818, 948} & $-1/2\sqrt{3}$
\\ \cline{2-9} & \multicolumn{3}{c|}{114, 152, 216, 272, 331, 364, 375, 441, 452, 636, 641} &
$-1/6$ & \multicolumn{3}{c|}{382, 678, 882, 985} & $i/2\sqrt{3}$
\\ \cline{6-9} & \multicolumn{3}{c|}{652, 761, 772, 831, 864, 875, 916, 934, 972, 046, 075} &
& \multicolumn{3}{c|}{328, 687, 828, 958} & $-i/2\sqrt{3}$
\\ \cline{2-9} & \multicolumn{3}{c|}{115, 124, 217, 226, 332, 347, 365, 424, 451, 615, 642} &
$i/6$ & \multicolumn{3}{c|}{234, 436} & $1/3$
\\ \cline{6-9} & \multicolumn{3}{c|}{673, 726, 771, 823, 856, 874, 953, 962, 971, 065, 074} &
& \multicolumn{3}{c|}{243, 463} & $-1/3$
\\ \cline{2-9} & \multicolumn{3}{c|}{142, 151, 262, 271, 323, 356, 374, 415, 442, 624, 637} &
$-i/6$ & \multicolumn{3}{c|}{253, 473, 512, 554, 567} & $i/3$
\\ \cline{6-9} & \multicolumn{3}{c|}{651, 717, 762, 832, 847, 865, 917, 926, 935, 047, 056} &
& \multicolumn{3}{c|}{235, 437, 521, 545, 576} & $-i/3$
\\ \cline{2-9} & \multicolumn{3}{c|}{583} & $1/\sqrt{3}$ & \multicolumn{3}{c|}{538} & $- 1/\sqrt{3}$
\\ \hline \hline $g_7$ & \multicolumn{3}{c|}{112, 143, 232, 245, 263, 315, 362, 448, 465, 619} & $1/3$ &
\multicolumn{3}{c|}{214, 333, 346, 412, 513, 518} & $2/3$
\\ & \multicolumn{3}{c|}{636, 665, 714, 768, 815, 844, 916, 945, 046, 069}
& & \multicolumn{3}{c|}{542, 549, 564, 566, 643, 869, 968} &
\\ \cline{2-9} & \multicolumn{3}{c|}{434, 939} &
$-1/3$ & \multicolumn{3}{c|}{838} & $-2/3$
\\ \cline{2-9} & \multicolumn{3}{c|}{372, 675, 724, 825, 854, 926, 955, 056, 079} &
$i/3$ & \multicolumn{3}{c|}{422, 523, 552, 574, 653, 978} & $2i/3$
\\ \cline{2-9} & \multicolumn{3}{c|}{122, 153, 255, 273, 325, 458, 475, 629, 778} &
$-i/3$ & \multicolumn{3}{c|}{224, 356, 528, 559, 576, 879} & $-2i/3$
\\ \cline{2-9} & \multicolumn{3}{c|}{131, 211, 341, 417, 640, 867, 960} & $1$ &
\multicolumn{3}{c|}{181, 282, 484, 787} & $1/\sqrt{3}$
\\ \cline{2-9} & \multicolumn{3}{c|}{737} & $-1$ &
\multicolumn{3}{c|}{686, 989} & $-1/\sqrt{3}$
\\ \cline{2-9} & \multicolumn{3}{c|}{221, 351, 877} & $i$ &
\multicolumn{3}{c|}{080} & $-2/\sqrt{3}$
\\ \cline{2-9} & \multicolumn{3}{c|}{427, 650, 970} & $-i$ &
\multicolumn{3}{c|}{} &
\\ \hline
\end{tabular}
\label{tab:couplings}
\end{center}
\end{table}

%
\section{Chiral multiplets/representations}
\label{sec:chiral_representation}
%

So far, we have performed classifications without explicitly
taking into account the left- and right-handed components of the
quark fields. However, it does not require great imagination to
see that the chiral properties are also conveniently studied in
that language, since chiral symmetry is defined as the symmetries
upon each chiral field. Hence, we define the left- and
right-handed (chiral or Weyl representation) quark fields as
\begin{eqnarray}
L \equiv q_L = { 1 - \gamma_5 \over 2 } q \mbox{,~~~~~and}~~~~~ R
\equiv q_R = { 1 + \gamma_5 \over 2 } q \, .
\end{eqnarray}
They form the fundamental representations of both the Lorentz group
and the chiral group,
\begin{eqnarray}
L: & & {\rm Lorentz:} ~~
(\frac12, 0)\, , ~~~~ {\rm Chiral:} ~~ (3,1) \, , \nonumber \\
\nonumber R: & & {\rm Lorentz:} ~~ (0, \frac12)\, , ~~~~ {\rm
Chiral:} ~~ (1,3)\, .
\end{eqnarray}
It is convenient first to note that $\gamma$-matrices are classified
into two categories; chiral-even and chiral-odd classes. The
chiral-even $\gamma$-matrices survive forming diquarks with
identical chiralities, while the chiral-odd ones form diquarks from
quarks with opposite chiralities. The chiral-even and -odd
$\gamma$-matrices are
\begin{eqnarray}
\mbox{chiral-even:} & & 1, \gamma_5, \sigma_{\mu \nu} \, , \nonumber \\
\nonumber \mbox{chiral-odd:} & & \gamma_\mu, \gamma_\mu \gamma_5 \,
.
\end{eqnarray}
Therefore, we have six non-vanishing diquarks in the chiral
representations,
\begin{eqnarray}
\left.
\begin{array}{l}
L^T C L = - L^T C \gamma_5 L \\
R^T C R = + R^T C \gamma_5 R
\end{array}
\right\} &~~~~~~& (0,0) \oplus (0,0)\, , ~~~~~ (\mathbf{\bar
3},~\mathbf{1}) \oplus (\mathbf{1},~\mathbf{\bar 3})\, ,
\nonumber \\
\left.
\begin{array}{l}
L^T C \gamma_\mu \gamma_5 R = + L^T C \gamma_\mu R  \\
R^T C \gamma_\mu \gamma_5 L = - R^T C \gamma_\mu L
\end{array}
\right\} &~~~~~~& (\frac12,\frac12) \oplus (\frac12, \frac12)\, ,
~~~~ (\mathbf{3},~\mathbf{3}) \oplus (\mathbf{3},~\mathbf{3})\, ,
\nonumber \\ \nonumber \left.
\begin{array}{l} L^T C \sigma_{\mu\nu} L \\ R^T C
\sigma_{\mu\nu} R \end{array} \right\} &~~~~~~& (1,0) \oplus (0,
1)\, , ~~~~~~~ (\mathbf{6},~\mathbf{1}) \oplus
(\mathbf{1},~\mathbf{6})\, ,
\end{eqnarray}
 where we have indicated the Lorentz and chiral representations of the diquarks.

For three quarks, we have
\begin{eqnarray}
(L + R)^3 \to \left\{ \begin{array}{l} LLL ~~~~ (\frac12,0) \oplus
(\frac32,0)\, , ~~~~ (\mathbf{1},\mathbf{1}) \oplus
(\mathbf{8},\mathbf{1}) \oplus
(\mathbf{8},\mathbf{1}) \oplus (\mathbf{10},\mathbf{1}) \\
LLR ~~~~ (0, \frac12) \oplus (1, \frac12)\, , ~~~~ (\mathbf{\bar
3},\mathbf{3}) \oplus (\mathbf{6},\mathbf{3})
\end{array}
\right.
\end{eqnarray}
and together with the terms where $L$ and $R$ are exchanged. Now we
discuss the independent fields in terms of the chiral
representations. Once again, for illustration we will discuss here
the case of the simplest Lorentz representation $(\frac12,0) \oplus
(0, \frac12)$ for the Dirac fields.

%
\subsection{Independent $(L L)L$ fields}
\label{subsec:LLL}
%

The $(L L) L$ must belong to one of the following chiral
representations: $(\mathbf{1},\mathbf{1}) \oplus
(\mathbf{8},\mathbf{1}) \oplus (\mathbf{8},\mathbf{1}) \oplus
(\mathbf{10},\mathbf{1})$. For each chiral representation, there
is one flavor representation available.

For $(\mathbf{1},~\mathbf{1}) \to \mathbf{1_f}$,
there are apparently two non-zero fields
\begin{eqnarray}
\begin{array}{l} \Lambda_{L1} = \epsilon_{abc}
\epsilon^{ABC} (L_{A}^{aT} C L_{B}^b) \gamma_5 L_{C}^c \, ,\\
\Lambda_{L2} = \epsilon_{abc} \epsilon^{ABC} (L_{A}^{aT} C \gamma_5
L_{B}^b) L_{C}^c \, ,\\
\Lambda_{L3} = \epsilon_{abc} \epsilon^{ABC} (L_{A}^{aT} C
\gamma_\mu \gamma_5 L_{B}^b) \gamma^\mu L_{C}^c = 0 \, ,
\end{array}
\end{eqnarray}
where $\Lambda^{L}_3$ vanishes because $\gamma_\mu \gamma_5$ is
chiral-odd
\begin{equation}\nonumber
L^T C \gamma_\mu \gamma_5 L = 0 \, .
\end{equation}
After performing the Fierz transformation to relate $\Lambda_{L i}$
and $\Lambda^\prime_{L i}$ as we have done before, and solving the
coupled equations, we find the solution that all such fields vanish.

For $(\mathbf{10},~\mathbf{1}) \to \mathbf{10_f}$, we would have
again two non-zero components:
\begin{eqnarray}
\begin{array}{l} \Delta^P_{L4} = \epsilon_{abc}
S_P^{ABC} (L_{A}^{aT} C \gamma_\mu L_{B}^b)
\gamma^\mu \gamma_5 L_{C}^c \, ,\\
\Delta^P_{L5} = \epsilon_{abc} S_P^{ABC} (L_{A}^{aT} C
\sigma_{\mu\nu} L_{B}^b) \sigma^{\mu\nu} \gamma_5 L_{C}^c \, .
\end{array}
\end{eqnarray}
Performing the Fierz transformation to relate $\Delta^P_{Li}$ and
$\Delta^{P\prime}_{Li}$, we obtain the solution that all such $(L
L) L$ fields vanish.

Finally for $(\mathbf{8},~\mathbf{1}) \to \mathbf{8_f}$, we may
consider once again two non-zero fields to start with
\begin{eqnarray}
\begin{array}{l} N^N_{L1} = \epsilon_{abc}
\epsilon^{ABD} \lambda_{DC}^N (L_{A}^{aT} C L_{B}^b) \gamma_5 L_{C}^c \, ,\\
N^N_{L2} = \epsilon_{abc} \epsilon^{ABD} \lambda_{DC}^N (L_{A}^{aT}
C \gamma_5 L_{B}^b) L_{C}^c \, .
\end{array}
\end{eqnarray}
Applying the Fierz transformation to relate $N^N_{Li}$ and
$N^{N\prime}_{Li}$, we obtain the solution
\begin{equation} \nonumber
N^N_{L2} = N^N_{L1} \, .
\end{equation}
Therefore, there is only one independent $(L L) L$ $\mathbf{8_f}$
field.

%
\subsection{Independent $(L L)R$ fields}
\label{subsec:LLR}
%
The chiral representations of $(L L)R$ are $(\mathbf{\bar
3},~\mathbf{3}) \oplus (\mathbf{6},~\mathbf{3})$. We will study them
separately in the following.

For
$(\mathbf{\bar 3},~\mathbf{3}) \to \mathbf{1_f}$,
there appears to exist two non-zero components among the five fields,
\begin{eqnarray}
\begin{array}{l} \Lambda_{M1} = \epsilon_{abc}
\epsilon^{ABC} (L_{A}^{aT} C L_{B}^b) \gamma_5 R_{C}^c \, ,\\
\Lambda_{M2} = \epsilon_{abc} \epsilon^{ABC} (L_{A}^{aT} C \gamma_5
L_{B}^b) R_{C}^c \, , \\
\Lambda_{M3} = \epsilon_{abc} \epsilon^{ABC} (L_{A}^{aT} C
\gamma_\mu
\gamma_5 L_{B}^b) \gamma^\mu R_{C}^c = 0 \, , \\
\Lambda_{M4} = \epsilon_{abc} \epsilon^{ABC} (L_{A}^{aT} C
\gamma_\mu
L_{B}^b) \gamma^\mu \gamma_5 R_{C}^c = 0 \, , \\
\Lambda_{M5} = \epsilon_{abc} \epsilon^{ABC} (L_{A}^{aT} C
\sigma_{\mu\nu} L_{B}^b) \sigma^{\mu\nu} \gamma_5 R_{C}^c = 0 \, ,
\end{array}
\end{eqnarray}
where $M$ (mixed) indicates that the fields contain both left and
right handed quarks. Performing the Fierz transformation to relate
$\Lambda_{Mi}$ and $\Lambda^\prime_{Mi}$, we obtain the following
relations
\begin{equation} \nonumber
\Lambda^{\prime}_{M4} = - \Lambda^{\prime}_{M3} = - 2 \Lambda_{M2} =
2 \Lambda_{M1} \, .
\end{equation}
We may consider other ten combinations formed by $(LR)$ and $(RL)$
diquarks, $(L R) L$ and $(R L) L$. However, they can be related to
the above ones of $(L L) R$ by a rearrangement of indices as well as
the Fierz transformation, for instance,
\begin{eqnarray}
\Lambda_{M6} = \epsilon_{abc} \epsilon^{ABC} (L_{A}^{aT} C R_{B}^b)
\gamma_5 L_{C}^c = \Lambda_{M1}^\prime \, .
\end{eqnarray}
Therefore, we have only one independent field.

For the chiral representation $(\mathbf{6},~\mathbf{3}) \rightarrow
\mathbf{10_f}$, we can write five fields containing diquarks formed
by five Dirac matrices. However, we can show that after performing
the Fierz transformation all fields vanish. Therefore, this
representation can not support three-quark fields.

The baryon fields of chiral representations $(\mathbf{\bar
3},~\mathbf{3}) \rightarrow \mathbf{8_f}$ can be formed
\begin{eqnarray}
\begin{array}{l} N^{N}_{M1} = \epsilon_{abc}
\epsilon^{ABD} \lambda_{DC}^N (L_A^{aT} C L_B^b) \gamma_5 R_C^c \, ,\\
N^{N}_{M2} = \epsilon_{abc} \epsilon^{ABD} \lambda_{DC}^N (L_A^{aT}
C \gamma_5
L_B^b) R_C^c \, , \\
N^{N}_{M3} = \epsilon_{abc} \epsilon^{ABD} \lambda_{DC}^N (L_A^{aT}
C \gamma_\mu
\gamma_5 L_B^b) \gamma^\mu R_C^c = 0 \, , \\
N^{N}_{M4} = \epsilon_{abc} \epsilon^{ABD} \lambda_{DC}^N (L_A^{aT}
C \gamma_\mu
L_B^b) \gamma^\mu \gamma_5 R_C^c = 0 \, , \\
N^{N}_{M5} = \epsilon_{abc} \epsilon^{ABD} \lambda_{DC}^N (L_A^{aT}
C \sigma_{\mu\nu} L_B^b) \sigma^{\mu\nu} \gamma_5 R_C^c = 0 \, ,
\end{array}
\label{NMfive}
\end{eqnarray}
where we see that there are two non-zero fields. Applying the Fierz
transformation, we can verify that there is only one independent
field with the following relations
\begin{equation}\nonumber
N^{N\prime}_{M4} = - N^{N\prime}_{M3} = - 2 N^{N}_{M2} = 2
N^{N}_{M1} \, .
\end{equation}

Another chiral representation $(\mathbf{6},~\mathbf{3}) \rightarrow
\mathbf{8_f}$ can be constructed by the combinations similar to
(\ref{NMfive}), for instance,
\begin{eqnarray}
N^{N}_{(6,3)1} = \epsilon_{abc} \epsilon^{ACD} \lambda_{DB}^N \{
(L_A^{aT} C L_B^b) \gamma_5 R_C^c + (L_B^{aT} C L_A^b) \gamma_5
R_C^c \} \, .
\end{eqnarray}
After similar algebra we can verify that all these fields vanish.

%
\subsection{A short summary of chiral representations}
\label{subsec:sum_chiral_rep}
%

To summarize this section, we find that possible chiral
representations for  Dirac spinor baryon fields without Lorentz
index are:
\begin{eqnarray}
\Lambda &=& \epsilon_{abc} \epsilon^{ABC} (L_A^{aT} C L_B^b)
\gamma_5 R_C^c + \epsilon_{abc} \epsilon^{ABC} (R_A^{aT} C R_B^b)
\gamma_5 L_C^c
\\ \nonumber &=&
\Lambda_{M1} + (L \leftrightarrow R) \, , \\ N^N_1 - N^N_2 &=& 2
\epsilon_{abc} \epsilon^{ABD} \lambda_{DC}^N (L_A^{aT} C L_B^b)
\gamma_5 R_C^c + 2 \epsilon_{abc} \epsilon^{ABD} \lambda_{DC}^N
(R_A^{aT} C R_B^b) \gamma_5 L_C^c
\\ \nonumber &=& 2 N^{N}_{M1} + (L \leftrightarrow R) \, ,
\\
N^N_1 + N^N_2 &=& 2 \epsilon_{abc} \epsilon^{ABD} \lambda_{DC}^N
(L_A^{aT} C L_B^b) \gamma_5 L_C^c + 2 \epsilon_{abc} \epsilon^{ABD}
\lambda_{DC}^N (R_A^{aT} C R_B^b) \gamma_5 R_C^c
\\ \nonumber &=& 2 N^{N}_{L1} + (L \leftrightarrow R) \, .
\end{eqnarray}
So we can see that the fields $\Lambda$ and $N^N_1 - N^N_2$ has a
type of $LLR \oplus RRL$, and belong to the chiral representation
$(\mathbf{\bar 3}, \mathbf{3}) \oplus (\mathbf{3}, \mathbf{\bar
3})$; while the field $N^N_1 + N^N_2$ has a type of $LLL \oplus
RRR$, and belongs to the chiral representation $(\mathbf{8},
\mathbf{1}) \oplus (\mathbf{1}, \mathbf{8})$.

The chiral properties of Rarita-Schwinger fields $\Big ({\rm
Lorentz~ rep.} (\mathbf{1}, \mathbf{1 \over 2})\oplus
(\mathbf{1\over2},\mathbf{1}) \Big )$ are listed in
Appendix~\ref{app:chiral_rep}. We summarize the results here:
\begin{eqnarray}
N^{N}_\mu &=& 2 \epsilon_{abc} \epsilon^{ABD} \lambda_{DC}^N
(L_A^{aT} C \gamma_\mu \gamma_5 R_B^b) \gamma_5 L_C^c + 2
\epsilon_{abc} \epsilon^{ABD} \lambda_{DC}^N (R_A^{aT} C \gamma_\mu
\gamma_5 L_B^b) \gamma_5 R_C^c \\ \nonumber &+& {1\over2}
\epsilon_{abc} \epsilon^{ABD} \lambda_{DC}^N (L_A^{aT} C L_B^b)
\gamma_\mu R_C^c + {1\over2} \epsilon_{abc} \epsilon^{ABD}
\lambda_{DC}^N (R_A^{aT} C R_B^b) \gamma_\mu L_C^c \, ,
\\
\Delta^P_{\mu} &=& 2 \epsilon_{abc} S_P^{ABC} (L_A^{aT} C \gamma_\mu
R_B^b) L_C^c + 2 \epsilon_{abc} S_P^{ABC} (R_A^{aT} C \gamma_\mu
L_B^b) R_C^c \, .
\end{eqnarray}
So we see that $N^{N}_\mu$ and $\Delta^P_{\mu}$ are of the type
$LLR \oplus RRL$, and belong to the chiral representation
$(\mathbf{6}, \mathbf{3}) \oplus (\mathbf{3}, \mathbf{6})$. The
(similar) results for $\Delta^P_{\mu\nu}$, which is of the type
$LLL \oplus RRR$, and belongs to the chiral representation
$(\mathbf{10}, \mathbf{1}) \oplus (\mathbf{1}, \mathbf{10})$, are
omitted here.

\section{Axial coupling constants}
\label{sec:axialcoupling}

As a simple application of the present mathematical formalism, we
can extract the (diagonal) axial coupling constants $g_A$ of
baryons. All information is contained in
Eqs.~(\ref{eq:U1Atransform}) and (\ref{eq:axialtransform}), from
which we can calculate the Abelian $U(1)_A$ axial coupling constant
$g^0_A$ and the non-Abelian $SU(3)_V \times SU(3)_A$ diagonal axial
coupling constants, $g^3_A$ and $g^8_A$, which can be extracted from
the chiral transformations $\delta_5$, $\delta_5^{b3}$ and
$\delta_5^{b8}$, respectively. The Abelian $g^0_A$ basically counts
the difference between the numbers of left- and right- handed quarks
in a baryon. In general, diagonal elements of the $SU(3)$ $g_A$'s
can be decomposed into $F$ and $D$ components, which are defined by
the axial vector current $A_\mu^a$ ($a = 0, 1, ... 8$)
\begin{eqnarray}
A^a_\mu = g_A^F {\rm tr} \bar {\mathfrak B}
\gamma_\mu \gamma_5 [\frac{\lambda_a}{2}, {\mathfrak B}] + g_A^D
{\rm tr} \bar { \mathfrak B} \gamma_\mu \gamma_5
\{\frac{\lambda_a}{2}, {\mathfrak B}\} \, ,
\end{eqnarray}
where $\mathfrak B$ is the $3 \times 3$ baryon matrix, Eq.
(\ref{eq:B}). Therefore, we have
\begin{eqnarray}
A^3_\mu &=&
(g_A^F + g_A^D) \Big ( p^+ p - n^+ n \Big ) \\ \nonumber &+& 2 g_A^F \Big (
(\Sigma^+)^+ \Sigma^+ - (\Sigma^-)^+ \Sigma^- \Big ) \\ \nonumber
&+& (g_A^F - g_A^D) \Big ( (\Xi^0)^+ \Xi^0 - (\Xi^-)^+ \Xi^- \Big ) \, ,
\\
A^8_\mu &=&
(\sqrt{3} g_A^F - {g_A^D \over \sqrt{3}})
\Big ( p^+ p + n^+ n \Big ) \\ \nonumber
&+&
{ 2 g_A^D \over \sqrt{3} }
\Big ( (\Sigma^+)^+ \Sigma^+ + (\Sigma^-)^+ \Sigma^- \Big ) \\
\nonumber &+& ( - \sqrt{3} g_A^F - {g_A^D \over \sqrt{3}}) \Big (
(\Xi^0)^+ \Xi^0 + (\Xi^-)^+ \Xi^- \Big ) - {2 g_A^D \over \sqrt{3}}
(\Lambda^8)^+ \Lambda^8 \, ,
\end{eqnarray}
where we omit the Lorentz part. In other words,
\begin{eqnarray}
g_A^3(N) &\sim& (g_A^F + g_A^D) {\bf I_z}\, , \; \; \; g_A^3(\Sigma)
\sim {2g_A^D} {\bf I_z} \, , \; \; \; g_A^3(\Xi) \sim (g_A^F -
g_A^D) {\bf I_z}\, ,
\\ \nonumber
g_A^8(N) &\sim& \sqrt{3} g_A^F - {g_A^D \over \sqrt{3}}\, , \; \; \;
g_A^8(\Sigma) \sim {2 g_A^D \over \sqrt{3}} \, , \; \; \; g_A^8(\Xi)
\sim - \sqrt{3} g_A^F - {g_A^D \over \sqrt{3}}\, , \; \; \;
g_A^8(\Lambda) \sim - {2 g_A^D \over \sqrt{3}} \, ,
\end{eqnarray}
for the octet parts. The operator $\bf I_z$ is the third component
of isospin. While the singlet part $g_A^0$ contains only the $D$
term and is trivial.

For the decuplet baryons, the $SU(3)$ coupling constants contain
only one $SU(3)$ irreducible term because the $SU(3)$ Clebsch-Gordan
series for $\mathbf{\bar {10}} \otimes \mathbf{10} \otimes
\mathbf{8}$ contains only one singlet. In order to extract the
coupling constants, we first rewrite Eqs.~(\ref{eq:U1Atransform})
and (\ref{eq:axialtransform}) in the following form, for all the
singlet, octet and decuplet baryon fields:
\begin{enumerate}

\item Because $\lambda^0_{11}=\lambda^0_{22}=\lambda^0_{33}$ for
$g^0_A$, the chiral transformations $\delta_5$ are identical for
all baryon fields within the same chiral representation, so we may
define $g_A^0$ by
\begin{eqnarray}\label{def:gA0}
\delta_5 B = i \gamma_5 { {\lambda^0}_{11} b_0\over 2} g_A^0 B = {i
\gamma_5 b_0 \over \sqrt{6}} g_A^0 B \, ,
\end{eqnarray}
where $B$ represents the baryon field, such as $\Lambda$ and $N_1^N
- N^N_2$ etc.

\item For $g^3_A$, because $\lambda^3_{11} = - \lambda^3_{22}$, the
chiral transformation $\delta_5^{b3}$ is proportional to the isospin
value of $\bf I_z$, which is related to $\lambda^3/2$. We factor it
out from the definition of $g_A^3$:
\begin{eqnarray}\label{def:gA3}
\delta^{b3}_5 B = i \gamma_5 b_3 g_A^3 {\bf I_z} B + \cdots \, ,
\end{eqnarray}
where dots $\cdots$ on the right hand side contain off-diagonal
terms.

\item For $g^8_A$, because $\lambda^8_{11} = \lambda^8_{22}$, the
chiral transformations $\delta_5^{b8}$ is the same for the baryon
fields belonging to one isospin multiplet. We define it to be
\begin{eqnarray}\label{def:gA8}
\delta^{b8}_5 B = i \gamma_5 { \bm{\lambda^8}_{11} b_8\over 2} g_A^8
B + \cdots = {i \gamma_5 b_8\over 2 \sqrt{3}} g_A^8 B + \cdots \, .
\end{eqnarray}

\end{enumerate}

\begin{table}[hbt]
\begin{center}
\caption{Axial coupling constants $g^0_A$, $g^3_A$ and $g^8_A$.
In the last column $\alpha =  {g_A^D /( g_A^F + g_A^D)}$.  }
\begin{tabular}{c | c | c | c | c | c | c}
\hline \hline $SU(3)_L \otimes SU(3)_R$ & $SU(3)_F$ & $\begin{array}{c} ~ \\ ~ \end{array}$ & $g_A^0$& $g_A^3$ & $g_A^8$ & ~~$\alpha$~~ \\
\hline & $\mathbf{1}$ & $\Lambda$ & -1 & -- & 0 & -- \\
\cline{2-7} & & $N_-$ & -1 & 1 & -1
\\ \cline{3-6} $(\mathbf{\bar 3},~\mathbf{3}) \oplus (\mathbf{3},~\mathbf{\bar 3})$ &
& $\Sigma_-$ & -1 & 0 & 2 &
\\ \cline{3-6} & $\mathbf{8}$ & $\Xi_-$ & -1 & -1 & -1 & 1
\\ \cline{3-6} & & $\Lambda_-$ & -1 & -- & -2 &
\\ \hline & & $N_+$ & 3 & 1 & 3 &
\\ \cline{3-6}  & & $\Sigma_+$ & 3 & 1 & 0 &
\\ \cline{3-6} $(\mathbf{8},~\mathbf{1}) \oplus (\mathbf{1},~\mathbf{8})$ & $\mathbf{8}$ & $\Xi_+$ & 3 &
1 & -3 & 0
\\ \cline{3-6} & & $\Lambda_+$ & 3 & -- & 0 &
\\ \hline & & $N_\mu$ & 1 & 5/3 & 1 &
\\ \cline{3-6}  & & $\Sigma_\mu$ & 1 & 2/3 & 2 &
\\ \cline{3-6}  & $\mathbf{8}$ & $\Xi_\mu$ & 1 & -1/3 & -3 &
3/5
\\ \cline{3-6}  & & $\Lambda_\mu$ & 1 & -- & -2 &
\\ \cline{2-7} $(\mathbf{3},~\mathbf{6}) \oplus (\mathbf{6},~\mathbf{3})$ & & $\Delta_\mu$
& 1 & 1/3 & 1 &
\\ \cline{3-6}  & & $\Sigma^*_\mu$ & 1 & 1/3 & 0 &
\\ \cline{3-6}  & $\mathbf{10}$ & $\Xi^*_\mu$ & 1 & 1/3 & -1 & --
\\ \cline{3-6} & & $\Omega_\mu$ & 1 & -- & -2 &
\\ \hline &  & $\Delta_{\mu\nu}$ & 3 & 1 & 3 &
\\ \cline{3-6}  & & $\Sigma^*_{\mu\nu}$ & 3 & 1 & 0 &
\\ \cline{3-6} $(\mathbf{10},~\mathbf{1}) \oplus (\mathbf{1},~\mathbf{10})$
& $\mathbf{10}$ & $\Xi^*_{\mu\nu}$ & 3 & 1 & -3 & --
\\ \cline{3-6} & & $\Omega_{\mu\nu}$ & 3 & -- & -6 &
\\ \hline
\end{tabular}
\label{tab:gA}
\end{center}
\end{table}

The resulting axial coupling constants $g^0_A$, $g^3_A$ and
$g^8_A$ are shown in Table~\ref{tab:gA}, where $\Lambda$ is the
(only) singlet field $\Lambda$; then $N_-$, $\Sigma_-$, $\Xi_-$
and $\Lambda_-$ are the octet fields of the type $N^N_1 - N^N_2$;
the $N_+$, $\Sigma_+$, $\Xi_+$ and $\Lambda_+$ are the octet
fields of the type $N^N_1 + N^N_2$; the $N_\mu$, $\Sigma_\mu$,
$\Xi_\mu$ and $\Lambda_\mu$ are the octet fields $N^N_\mu$; the
$\Delta_\mu$, $\Sigma^*_\mu$, $\Xi^*_\mu$ and $\Omega_\mu$ are the
decuplet fields $\Delta^P_\mu$; $\Delta_{\mu\nu}$,
$\Sigma^*_{\mu\nu}$, $\Xi^*_{\mu\nu}$ and $\Omega_{\mu\nu}$ are
the decuplet fields $\Delta^P_{\mu\nu}$.

From the values in Table~\ref{tab:gA}, one can compute the $F$ and
$D$ couplings easily. The resulting $F/D$ ratio,
\begin{eqnarray}
\alpha = {g_A^D \over g_A^F + g_A^D} \, ,
\end{eqnarray}
is also tabulated in the last column of Table~\ref{tab:gA}.
Empirically, $\alpha \sim 0.6$, which is fairly close to the $SU(6)$
quark model value. In the present formalism we see that only the
$(\mathbf{3},~\mathbf{6}) \oplus (\mathbf{6},~\mathbf{3})$ chiral
multiplet/representation reproduces this value. Previous works have
shown that this value is physically related to the coupling of the
nucleon to the $\Delta(1232)$, as demonstrated in the
Adler-Weisberger sum rule~\cite{adler,weissberger}. This was also
shown algebraically by Weinberg~\cite{Weinberg:1969hw}. In both
cases, saturation of the pion (axial-vector) induced transition from
the nucleon to the $\Delta(1232)$ is
essential~\cite{Donoghue:1977yh}. In the present study, this is
realized by the chiral representation which includes both the
nucleon (isospin 1/2) and delta (isospin 3/2) states.

It is also interesting that Table~\ref{tab:gA} shows that
$g_A^3(N) = 5/3, g_A^0(N) = 1$ for $(\mathbf{3},~\mathbf{6})
\oplus (\mathbf{6},~\mathbf{3})$, while $g_A^3(N) = 1, g_A^0(N) =
-1$ for $(\mathbf{\bar 3},~\mathbf{3}) \oplus
(\mathbf{3},~\mathbf{\bar 3})$. $g_A^0$ corresponds to the
so-called nucleon spin value, as measured in polarized deep
inelastic scattering. A suitable superposition of the two chiral
representations may improve the nucleon axial coupling in either
the isovector and/or isosinglet sectors. The importance of such
mixing for the isovector axial coupling constant  has been
emphasized by Weinberg since the late 1960-s,
Ref.~\cite{Weinberg:1969hw}. Here we have found the same result
for the isovector, as well as extended it to the isosinglet part
in a purely algebraic manner.

%
\section{Summary}
\label{sec:summary}

In this paper we have performed a complete classification of flavor
vector and chiral symmetries, and established several types of
independent relativistic $SU(3)$ baryon interpolating fields. The
three-quark fields may belong to one of several different Lorentz
group representations which fact imposes certain constraints on
possible chiral symmetry representations. This is due to the Pauli
principle and has been explicitly verified by the method of Fierz
transformations. As the present results reflect essentially the
Pauli principle, they can be conveniently summarized as shown in
Table~\ref{tab:summary} by using the permutation symmetry group
properties/representations. This table ``explains" also the previous
results for the case of isospin $SU(2)_L \times
SU(2)_R$~\cite{Nagata:2007di}. From this table we have explicated
the effective role of the Pauli principle in separate sectors of the
left- and right-handed fermions.

\begin{table}[tbh]
\begin{center}
\caption{Structure of allowed three-quark baryon fields.}
\begin{tabular}{c | c | c | c | c | c}
\hline \hline Lorentz & J=Spin & $\begin{array}{c} \mbox{Young table} \\
\mbox{for Chiral rep.} \end{array}$
& Chiral $SU(2)$ & Chiral $SU(3)$  & Flavor $SU(3)$ \\
\hline $(\frac12,0)\oplus(0,\frac12)$ & $1/2$ & $\begin{array}{c}
([21],-)\oplus(-,[21]) \\ ([1],[11])\oplus([11],[1]) \end{array}$
& $(\frac12,0)\oplus(0,\frac12)$ 
& $ \begin{array}{c} (8,1)\oplus(1,8) \\
(3,\bar{3})\oplus(\bar{3},3) \end{array}$
& $ \begin{array}{c} 8 \\ 1, 8 \end{array}$ \\
\hline
$(1,\frac12)\oplus(\frac12,1)$ & 
$3/2$ &  $([2],[1])\oplus([1],[2])$ & $(1,\frac12)\oplus(\frac12,1)$ 
& $(6,3)\oplus(3,6)$  & $8$, $10$ \\
\hline
$(\frac32,0)\oplus(0,\frac32)$ & $3/2$ & $([3],-)\oplus(-,[3])$ &
$(\frac32,0)\oplus(0,\frac32)$ 
& $(10,1)\oplus(1,10)$  & $10$ \\
\hline
\end{tabular}
\label{tab:summary}
\end{center}
\end{table}
In the real world, with spontaneous breaking of chiral symmetry,
physical states of pure chiral (axial) symmetry representation do
not occur, but in general they can mix in a state having a definite
flavor symmetry. The present results show that the three-quark
structures accommodate only a few (sometimes just one) chiral
representations, for instance, for the total spin 1/2 field of Dirac
spinor, the allowed chiral representations are two having the
structure of Young tableaux $([21],-)$ and ([1],[11]), where $-$
indicates singlet. The $([21],-)$ representation corresponds
respectively to $(\frac12,0)$ and $(8,1)$ of $SU(2)$ and $SU(3)$,
whereas the $([1],[11])$ corresponds to $(\frac12,0)$ and
$(3,\bar{3})$ of $SU(2)$ and $SU(3)$, respectively. Note that the
chiral representations have the same structure as those of the
Lorentz group. In this way, the Lorentz (spin) and flavor structures
are combined into a structure of total permutation symmetry. As
shown in the computation of $g_A$, in general, various couplings
depend on the chiral representations with possible mixing. Such
comparison may be useful for further understanding of the internal
structure of hadrons in relation to chiral symmetry.

\section*{Acknowledgments}
%

H.X.C. is grateful to Monkasho for their support of his stay at the
Research Center for Nuclear Physics where this work was done. V.D
and K.N thank Prof H.~Toki for his hospitality during their stay at
RCNP. A.H. is supported in part by the Grant for Scientific Research
((C) No.19540297) from the Ministry of Education, Culture, Science
and Technology, Japan. K.N is supported by the National Science
Council (NSC) of Republic of China under grant
No.~NSC96-2119-M-002-001. S.L.Z. was supported by the National
Natural Science Foundation of China under Grants 10625521 and
10721063 and Ministry of Education of China.

%
\appendix
%
%
\section{Rarita-Schwinger fields}
\label{app:baryon1}
%

In this appendix, we study the properties of Rarita-Schwinger
fields, in the form of
\begin{equation}
B_\mu(x) \sim \epsilon_{abc} (q^{aT}_A (x) C \Gamma_1 q^b_B (x))
\Gamma_2 q^c_C (x) \, ,
\end{equation}
where there are eight possible pairs of $\Gamma_1$ and $\Gamma_2$,
\begin{eqnarray}
\label{eq:rariltordering} (\Gamma_1,~\Gamma_2) &=&
(\mathbf{1},~\gamma_\mu),~ (\gamma_5,~\gamma_{\mu}\gamma_5),~
(\gamma_{\mu}\gamma_5,~\gamma_5),~
(\gamma^\nu\gamma_5,~\sigma_{\mu\nu}\gamma_5),~
\\ \nonumber && (\gamma_\mu,~\mathbf{1}),~
(\gamma^\nu,~\sigma_{\mu\nu}),~ (\sigma_{\mu\nu},~\gamma^\nu),~
(\sigma_{\mu\nu}\gamma_5,~\gamma^\nu\gamma_5).
\end{eqnarray}
The fields formed by these $(\Gamma_1,\Gamma_2)$ pairs are labeled
by the subscript $i=(1,\cdots,8)$ with the ordering of
Eq.~(\ref{eq:rariltordering}). The discussion is separated into
singlet, decuplet and octet cases.

\subsubsection{Flavor singlet baryon}

For flavor singlet fields, there are four apparently non-zero fields
\begin{eqnarray}
\begin{array}{l}
\Lambda_{1\mu} = \epsilon_{abc} \epsilon^{ABC} (q_A^{aT} C q_B^b)
\gamma_\mu q_C^c \, ,
\\ \Lambda_{2\mu} = \epsilon_{abc} \epsilon^{ABC} (q_A^{aT} C \gamma_5 q_B^b) \gamma_\mu \gamma_5
q_C^c \, ,
\\ \Lambda_{3\mu} = \epsilon_{abc} \epsilon^{ABC} (q_A^{aT} C \gamma_\mu \gamma_5 q_B^b) \gamma_5
q_C^c \, ,
\\ \Lambda_{4\mu} = \epsilon_{abc} \epsilon^{ABC} (q_A^{aT} C \gamma^\nu \gamma_5 q_B^b) \sigma_{\mu\nu}
\gamma_5 q_C^c \, .
\end{array}
\end{eqnarray}
As before, the Fierz transformed fields (primed fields) are just the
corresponding unprimed ones, $\Lambda^\prime_{i\mu} = \Lambda_{i\mu}
$. On the other hand, by applying the Fierz rearrangement (see
Appendix.~\ref{app:fierz}), we obtain four equations
\begin{eqnarray}\nonumber
\begin{array}{l}
\Lambda_{1\mu} = - {1 \over 4} \Lambda_{1\mu}^\prime - {1 \over 4}
\Lambda_{2\mu}^\prime + {1 \over 4} \Lambda_{3\mu}^\prime - {i \over
4} \Lambda_{4\mu}^\prime \, , \\ \Lambda_{2\mu} = - {1 \over 4}
\Lambda_{1\mu}^\prime - {1 \over 4} \Lambda_{2\mu}^\prime - {1 \over
4} \Lambda_{3\mu}^\prime + {i \over 4} \Lambda_{4\mu}^\prime \, , \\
\Lambda_{3\mu} = {1 \over 4} \Lambda_{1\mu}^\prime - {1 \over 4}
\Lambda_{2\mu}^\prime - {1 \over
4} \Lambda_{3\mu}^\prime - {i \over 4} \Lambda_{4\mu}^\prime \, , \\
\Lambda_{4\mu} = {3i \over 4} \Lambda_{1\mu}^\prime - {3i \over 4}
\Lambda_{2\mu}^\prime + {3i \over 4} \Lambda_{3\mu}^\prime + {1
\over 4} \Lambda_{4\mu}^\prime \, .
\end{array}
\end{eqnarray}
Thus we find the following solution
\begin{equation} \nonumber
\Lambda_{1\mu} = - \Lambda_{2\mu} = \Lambda_{3\mu} = - {i \over 3}
\Lambda_{4\mu} = \gamma_\mu \gamma_5 \Lambda_1 \, , ~~~~~
\Lambda_{6\mu} = \Lambda_{7\mu} = \Lambda_{8\mu} = 0.
\end{equation}
We see that there is only one non-vanishing independent field.
However, it has a structure of $\gamma_\mu \Lambda_i$\, . Therefore,
they are all Dirac fields, and there is no flavor singlet fields of
the Rarita-Schwinger type.

\subsubsection{Flavor decuplet baryon}

For flavour decuplet fields, we have four potentially non-zero
interpolators
\begin{eqnarray}
\begin{array}{l}
\Delta^P_{5\mu} = \epsilon_{abc} S_P^{ABC} (q_A^{aT} C \gamma_\mu
q_B^b) q_C^c \, ,
\\ \Delta^P_{6\mu} = \epsilon_{abc} S_P^{ABC} (q_A^{aT} C \gamma^\nu q_B^b) \sigma_{\mu\nu} q_C^c \, ,
\\ \Delta^P_{7\mu} = \epsilon_{abc} S_P^{ABC} (q_A^{aT} C \sigma_{\mu\nu} q_B^b) \gamma^\nu q_C^c \, ,
\\ \Delta^P_{8\mu} = \epsilon_{abc} S_P^{ABC} (q_A^{aT} C \sigma_{\mu\nu} \gamma_5 q_B^b) \gamma^\nu \gamma_5
q_C^c \, .
\end{array}
\end{eqnarray}
As before, the Fierz transformed fields can be related to the
corresponding unprimed ones, $\Delta^{P\prime}_{i\mu} = -
\Delta^P_{i\mu}$. On the other hand, by applying the Fierz
rearrangement to relate $\Delta_{i\mu}^N$ and
$\Delta_{i\mu}^{N\prime}$, we obtain the solution
\begin{equation} \nonumber
\Delta^P_{5\mu} = i \Delta^P_{6\mu} = - i \Delta^P_{7\mu} = i
\Delta^P_{8\mu} \, .
\end{equation}
There are no Dirac decuplet fields. Therefore, we obtain one extra
non-vanishing field.

\subsubsection{Flavor octet baryon}

To study the octet baryon fields, we start with eight baryon fields:
\begin{eqnarray}
\begin{array}{l}
N^N_{1\mu} = \epsilon_{abc} \epsilon^{ABD} \lambda_{DC}^N (q_A^{aT}
C q_B^b) \gamma_\mu q_C^c \, ,
\\ N^N_{2\mu} = \epsilon_{abc}
\epsilon^{ABD} \lambda_{DC}^N (q_A^{aT} C \gamma_5 q_B^b) \gamma_\mu
\gamma_5 q_C^c \, ,
\\ N^N_{3\mu} = \epsilon_{abc}
\epsilon^{ABD} \lambda_{DC}^N (q_A^{aT} C \gamma_\mu \gamma_5 q_B^b)
\gamma_5 q_C^c \, ,
\\ N^N_{4\mu} = \epsilon_{abc}
\epsilon^{ABD} \lambda_{DC}^N (q_A^{aT} C \gamma^\nu \gamma_5 q_B^b)
\sigma_{\mu\nu} \gamma_5 q_C^c \, ,
\\ N^N_{5\mu} =
\epsilon_{abc} \epsilon^{ABD} \lambda_{DC}^N (q_A^{aT} C \gamma_\mu
q_B^b) q_C^c = 0  \, ,
\\ N^N_{6\mu} = \epsilon_{abc}
\epsilon^{ABD} \lambda_{DC}^N (q_A^{aT} C \gamma^\nu q_B^b)
\sigma_{\mu\nu} q_C^c = 0 \, ,
\\ N^N_{7\mu} = \epsilon_{abc}
\epsilon^{ABD} \lambda_{DC}^N (q_A^{aT} C \sigma_{\mu\nu} q_B^b)
\gamma^\nu q_C^c = 0 \, ,
\\ N^N_{8\mu} = \epsilon_{abc}
\epsilon^{ABD} \lambda_{DC}^N (q_A^{aT} C \sigma_{\mu\nu} \gamma_5
q_B^b) \gamma^\nu \gamma_5 q_C^c = 0 \, .
\end{array}
\end{eqnarray}
There are four zero fields, but the corresponding Fierz transformed
ones are non-zero. By using the Jacobi identity in
Eq.~(\ref{eq:Jacobi}), we obtain
\begin{eqnarray} \nonumber
N_{1\mu}^{N\prime} = - {1 \over 2} N_{1\mu}^N \, ,
N_{2\mu}^{N\prime} = - {1 \over 2} N_{2\mu}^N \, ,
N_{3\mu}^{N\prime} = - {1 \over 2} N_{3\mu}^N \, ,
N_{4\mu}^{N\prime} = - {1 \over 2} N_{4\mu}^N \, .
\end{eqnarray}
Similarly, performing the Fierz transformation to relate
$N_{i\mu}^N$ and $N_{i\mu}^{N\prime}$, we obtain the solution
\begin{eqnarray}
\begin{array}{l}
N_{4\mu}^N = - i N_{1\mu}^N + i N_{2\mu}^N - i N_{3\mu}^N \, ,
\\ N_{5\mu}^{N\prime} = - {1 \over 2} N_{1\mu}^N + {1 \over 2} N_{2\mu}^N - {1 \over 2} N_{3\mu}^N \, ,
\\ N_{6\mu}^{N\prime} = - i N_{1\mu}^N + i N_{2\mu}^N + {i \over 2} N_{3\mu}^N \, ,
\\ N_{7\mu}^{N\prime} = i N_{1\mu}^N + {i \over 2} N_{2\mu}^N + i N_{3\mu}^N \, ,
\\ N_{8\mu}^{N\prime} = {i \over 2} N_{1\mu}^N + i N_{2\mu}^N - i N_{3\mu}^N \,
.
\end{array}
\end{eqnarray}
Thus we have shown that there are three different kinds of octets.
However, $N_{1\mu}^N$ and $N_{2\mu}^N$ are nothing but $\gamma_\mu
\gamma_5 N_1^N$ and $\gamma_\mu \gamma_5 N_2^N$ (see
Eqs.~(\ref{eq:fiveNs})). Therefore, we only obtain one extra octet
baryon field. It is formed by using the projection operator:
\begin{eqnarray} \nonumber
P^{3/2}_{\mu\nu} = (g_{\mu\nu} - {1\over4}\gamma_\mu\gamma_\nu) \, ,
\end{eqnarray}
as
\begin{eqnarray} \nonumber
N^N_\mu = P^{3/2}_{\mu\nu} N^{N}_{3\nu} &=& (g_{\mu\nu} -
{1\over4}\gamma_\mu\gamma_\nu) \epsilon_{abc} \epsilon^{ABD}
\lambda_{DC}^N (q_A^{aT} C \gamma_\mu \gamma_5 q_B^b) \gamma_5 q_C^c
\\ \nonumber &=& N^N_{3\mu} + {1\over4} \gamma_\mu \gamma_5 (N^N_1 -
N^N_2) \, .
\end{eqnarray}

%
\section{Tensor Fields}
\label{app:baryon2}
%

In this appendix, we study the antisymmetric tensor baryons fields
$J_{\mu\nu}$ with $J_{\mu\nu} = - J_{\nu\mu}$. For the tensor
fields, we can form nine three-quark fields where the possible pairs
of $\Gamma_1$ and $\Gamma_2$ are
\begin{eqnarray}\label{eq:tensorordering}
(\Gamma_1,~\Gamma_2) &=& (\gamma_\mu,~\gamma_\nu\gamma_5) - (\mu
\leftrightarrow \nu),~ (\gamma_\mu\gamma_5,~\gamma_\nu) - (\mu
\leftrightarrow \nu),~ \\ \nonumber && \epsilon_{\mu\nu\rho\sigma}
(\gamma^\rho,~\gamma^\sigma),~ \epsilon_{\mu\nu\rho\sigma}
(\gamma^\rho\gamma_5,~\gamma^\sigma\gamma_5),~
(\mathbf{1},~\sigma_{\mu\nu}\gamma_5),~
(\gamma_5,~\sigma_{\mu\nu}),~ \\ \nonumber &&
(\sigma_{\mu\nu},~\gamma_5),~
(\sigma_{\mu\nu}\gamma_5,~\mathbf{1}),~ \epsilon_{\mu\nu\rho\sigma}
(\sigma_{\rho l}, \sigma_{\sigma l}) \, .
\end{eqnarray}
The fields formed by these $(\Gamma_1,\Gamma_2)$ pairs are labeled
by the subscript $i=(1,\cdots,9)$ with the ordering of
Eq.~(\ref{eq:tensorordering}). The discussion is separated into
singlet, decuplet and octet cases.

\subsubsection{Flavor singlet baryon}
The flavour singlet baryon fields have four potentially non-zero
interpolators among nine fields:
\begin{eqnarray}
\begin{array}{l} \Lambda_{2\mu\nu} = \epsilon_{abc} \epsilon^{ABC}
(q_A^{aT} C \gamma_{\mu} \gamma_5 q_B^b) \gamma_\nu q_C^c - ( \mu
\leftrightarrow \nu ) \, ,
\\ \Lambda_{4\mu\nu} = \epsilon_{abc} \epsilon^{ABC} \epsilon_{\mu\nu\rho\sigma}
(q_A^{aT} C \gamma_{\rho} \gamma_5 q_B^b) \gamma_{\sigma} \gamma_5
q_C^c \, ,
\\ \Lambda_{5\mu\nu} = \epsilon_{abc} \epsilon^{ABC} (q_A^{aT} C q_B^b) \sigma_{\mu\nu} \gamma_5
q_C^c \, ,
\\ \Lambda_{6\mu\nu} = \epsilon_{abc} \epsilon^{ABC} (q_A^{aT} C \gamma_5 q_B^b) \sigma_{\mu\nu} q_C^c
\, .
\end{array}
\end{eqnarray}
As before, the Fierz transformed fields are just the corresponding
unprimed ones, $\Lambda^\prime_{i\mu\nu} = \Lambda_{i\mu\nu} $. On
the other hand, by applying the Fierz rearrangement to relate
$\Lambda_{i\mu\nu}$ and $\Lambda_{\i\mu\nu}^{\prime}$, we obtain the
solution:
\begin{eqnarray} \nonumber
i \Lambda_{2\mu\nu} = \Lambda_{4\mu\nu} = 2 \Lambda_{5\mu\nu} = - 2
\Lambda_{6\mu\nu} \, .
\end{eqnarray}
Therefore, there is only one independent field. However, it has a
structure of $\sigma_{\mu\nu} \Lambda_i$\, . Therefore, there are no
extra fields.

\subsubsection{Flavor decuplet baryon}
The flavour decuplet baryon fields have five potentially non-zero
interpolators:
\begin{eqnarray}
\begin{array}{l}
\Delta^P_{1\mu\nu} = \epsilon_{abc} S^{ABC} (q_A^{aT} C \gamma_{\mu}
q_B^b) \gamma_\nu \gamma_5 q_C^c - ( \mu \leftrightarrow \nu ) \, ,
\\ \Delta^P_{3\mu\nu} = \epsilon_{abc} S^{ABC} \epsilon_{\mu\nu\rho\sigma}
(q_A^{aT} C \gamma_{\rho} q_B^b) \gamma_{\sigma} q_C^c  \, ,
\\ \Delta^P_{7\mu\nu} = \epsilon_{abc} S^{ABC} (q_A^{aT} C \sigma_{\mu\nu} q_B^b) \gamma_5
q_C^c \, ,
\\ \Delta^P_{8\mu\nu} = \epsilon_{abc} S^{ABC} (q_A^{aT} C \sigma_{\mu\nu} \gamma_5 q_B^b) q_C^c \, ,
\\ \Delta^P_{9\mu\nu} = \epsilon_{abc} S^{ABC} \epsilon_{\mu\nu\rho\sigma}
(q_A^{aT} C \sigma_{\rho l} q_B^b) \sigma_{\sigma l} q_C^c  \, .
\end{array}
\end{eqnarray}
As before, the Fierz transformed fields can be related to the
corresponding unprimed ones, $\Delta^{P\prime}_{i\mu\mu} = -
\Delta^P_{i\mu\mu}$. On the other hand, by applying the Fierz
rearrangement to relate $\Delta^P_{i\mu\nu}$ and
$\Delta_{i\mu\nu}^{P\prime}$, we obtain two independent fields:
$\Delta^P_{1\mu\nu}$ and $\Delta^P_{7\mu\nu}$:
\begin{eqnarray} \nonumber
\Delta^P_{3\mu\nu} = - i \Delta^P_{1\mu\nu} \, , \Delta^P_{8\mu\nu}
= i \Delta^P_{1\mu\nu} + \Delta^P_{7\mu\nu} \, , \Delta^P_{9\mu\nu}
= - i \Delta^P_{1\mu\nu} - 2 \Delta^P_{7\mu\nu} \, .
\end{eqnarray}
The first one $\Delta^P_{1\mu\nu}$ can be related to the
Rarita-Schwinger baryon fields, but the second one
$\Delta^P_{7\mu\nu}$ can not. Therefore, we obtain one extra
decuplet fields. It is formed by using the projection operator:
\begin{eqnarray} \nonumber \Gamma^{\mu\nu\alpha\beta} = (
g^{\mu\alpha}g^{\nu\beta} -
{1\over2}g^{\nu\beta}\gamma^{\mu}\gamma^\alpha +
{1\over2}g^{\mu\beta}\gamma^\nu\gamma^\alpha +
{1\over6}\sigma^{\mu\nu}\sigma^{\alpha\beta})\, ,
\end{eqnarray}
as
\begin{eqnarray}
\nonumber \Delta^P_{\mu\nu} = \Gamma^{\mu\nu\alpha\beta}
\Delta^P_{7\mu\nu} &=& \Gamma^{\mu\nu\alpha\beta} \epsilon_{abc}
S^{ABC} (q_A^{aT} C \sigma_{\mu\nu} q_B^b) \gamma_5 q_C^c
\\ \nonumber &=& \Delta^P_{7\mu\nu} - {i\over2} \gamma_\mu \gamma_5
\Delta^P_{5\nu} + {i\over2} \gamma_\nu \gamma_5 \Delta^P_{5\mu} \, .
\end{eqnarray}

\subsubsection{Flavor octet baryon}
To study the octet baryon fields, we start with nine octet baryon
fields
\begin{eqnarray}
\begin{array}{l}
N^N_{1\mu\nu} = \epsilon_{abc} \epsilon^{ABD} \lambda_{DC}^N
(q_A^{aT} C \gamma_{\mu} q_B^b) \gamma_\nu \gamma_5 q_C^c - ( \mu
\leftrightarrow \nu ) = 0 \, ,
\\ N^N_{2\mu\nu} = \epsilon_{abc} \epsilon^{ABD} \lambda_{DC}^N (q_A^{aT} C \gamma_{\mu} \gamma_5
q_B^b) \gamma_\nu q_C^c - ( \mu \leftrightarrow \nu ) \, ,
\\ N^N_{3\mu\nu} = \epsilon_{abc} \epsilon^{ABD} \lambda_{DC}^N \epsilon_{\mu\nu\rho\sigma}
(q_A^{aT} C \gamma_{\rho} q_B^b) \gamma_{\sigma} q_C^c = 0  \, ,
\\ N^N_{4\mu\nu} = \epsilon_{abc} \epsilon^{ABD} \lambda_{DC}^N \epsilon_{\mu\nu\rho\sigma}
(q_A^{aT} C \gamma_{\rho} \gamma_5 q_B^b) \gamma_{\sigma} \gamma_5
q_C^c \, ,
\\ N^N_{5\mu\nu} = \epsilon_{abc} \epsilon^{ABD} \lambda_{DC}^N (q_A^{aT} C q_B^b) \sigma_{\mu\nu} \gamma_5
q_C^c \, ,
\\ N^N_{6\mu\nu} = \epsilon_{abc} \epsilon^{ABD} \lambda_{DC}^N (q_A^{aT} C \gamma_5 q_B^b) \sigma_{\mu\nu} q_C^c \, ,
\\ N^N_{7\mu\nu} = \epsilon_{abc} \epsilon^{ABD} \lambda_{DC}^N (q_A^{aT} C \sigma_{\mu\nu} q_B^b) \gamma_5
q_C^c = 0 \, ,
\\ N^N_{8\mu\nu} = \epsilon_{abc} \epsilon^{ABD}
\lambda_{DC}^N (q_A^{aT} C \sigma_{\mu\nu} \gamma_5 q_B^b) q_C^c = 0
\, ,
\\ N^N_{9\mu\nu} = \epsilon_{abc} \epsilon^{ABD} \lambda_{DC}^N \epsilon_{\mu\nu\rho\sigma}
(q_A^{aT} C \sigma_{\rho l} q_B^b) \sigma_{\sigma l} q_C^c = 0  \, .
\end{array}
\end{eqnarray}
There are five zero fields, but the Fierz transformed ones are
non-zero. By using the Jacobi identity in Eq.~(\ref{eq:Jacobi}), we
obtain
\begin{eqnarray} \nonumber
N^{N\prime}_{2\mu\nu} = - {1\over2} N^N_{2\mu\nu}\, ,
N^{N\prime}_{4\mu\nu} = - {1\over2} N^N_{4\mu\nu}\, ,
N^{N\prime}_{5\mu\nu} = - {1\over2} N^N_{5\mu\nu}\, ,
N^{N\prime}_{6\mu\nu} = - {1\over2} N^N_{6\mu\nu}\, .
\end{eqnarray}
Similarly, performing the Fierz transformation to relate
$N^N_{i\mu\nu}$ and $N_{i\mu\nu}^{N\prime}$, we find that there are
three independent fields $N_{2\mu\nu}^{N}$, $N_{5\mu\nu}^{N}$ and
$N_{6\mu\nu}^{N}$. Here are the relations:
\begin{eqnarray}
\begin{array}{l} N^N_{4\mu\nu} = - i N^N_{2\mu\nu} - N^N_{5\mu\nu} +
N^N_{6\mu\nu} \, ,
\\ N^{N\prime}_{1\mu\nu} = - {1\over2} N^N_{2\mu\nu} + i N^N_{5\mu\nu} - i N^N_{6\mu\nu}
\, ,
\\ N^{N\prime}_{3\mu\nu} = {i\over2} N^N_{2\mu\nu} - {1\over2} N^N_{5\mu\nu} + {1\over2} N^N_{6\mu\nu}
\, ,
\\ N^{N\prime}_{7\mu\nu} = - {i\over2} N^N_{2\mu\nu} - {1\over2} N^N_{5\mu\nu}
\, ,
\\ N^{N\prime}_{8\mu\nu} = {i\over2} N^N_{2\mu\nu} - {1\over2} N^N_{6\mu\nu}
\, ,
\\ N^{N\prime}_{9\mu\nu} = - N^N_{5\mu\nu} - N^N_{6\mu\nu}
\, .
\end{array}
\end{eqnarray}
All these three fields can be related to the Dirac spinor and
Rarita-Schwinger fields. Therefore, there are no extra octet fields.

%
\section{Fierz Transformation}
\label{app:fierz}
%

In this appendix, we list the Fierz transformations used in our
calculation, which are proved by using
$Mathematica$~\cite{Maruhn:2000af}. Here we would like to show only
the change in the structure of Lorentz indices of direct products of
two Dirac matrices under the Fierz rearrangement. Therefore, in the
following equations, we do not include the minus sign which arises
from the exchange of quark fields. The formulae go for the three
cases corresponding to the Dirac, Rarita-Schwinger and tensor fields
when applied to three-quark fields.

\begin{enumerate}

\item
Products of two Dirac matrices without Lorentz indices:
\begin{eqnarray}
\left (
\begin{array}{l}
\mathbf{1} \otimes \gamma_5
\\ \gamma_\mu \otimes \gamma^\mu  \gamma_5
\\ \sigma_{\mu\nu} \otimes \sigma^{\mu\nu} \gamma_5
\\ \gamma_{\mu} \gamma_5 \otimes \gamma^{\mu}
\\ \gamma_5 \otimes \mathbf{1}
\end{array} \right )_{a b, c d}
= \left (
\begin{array}{lllll}
{1\over4} & - {1\over4} & {1\over8} & {1\over4} & {1\over4}
\\ - 1 & - {1\over2} & 0 & - {1\over2} & 1
\\ 3 & 0 & -{1\over2} & 0 & 3
\\ 1 & -{1\over2} & 0 & - {1\over2} & - 1
\\ {1\over4} & {1\over4} & {1\over8} & - {1\over4} & {1\over4}
\end{array} \right )
\left (
\begin{array}{l}
\mathbf{1} \otimes \gamma_5
\\ \gamma_\mu \otimes \gamma^\mu  \gamma_5
\\ \sigma_{\mu\nu} \otimes \sigma^{\mu\nu} \gamma_5
\\ \gamma_{\mu} \gamma_5 \otimes \gamma^{\mu}
\\ \gamma_5 \otimes \mathbf{1}
\end{array} \right )_{a d, b c}
\label{Fierz5}
\end{eqnarray}

\item
Products of two Dirac matrices with one Lorentz index:
\begin{eqnarray}
\left (
\begin{array}{l}
\mathbf{1} \otimes \gamma^\mu
\\ \gamma^\mu \otimes \mathbf{1}
\\ \gamma_5 \otimes \gamma_\mu \gamma_5
\\ \gamma_\mu \gamma_5 \otimes \gamma_5
\\ \gamma^{\nu} \otimes \sigma_{\mu\nu}
\\ \sigma_{\mu\nu} \otimes \gamma^{\nu}
\\ \gamma^{\nu} \gamma_5 \otimes \sigma_{\mu\nu} \gamma_5
\\ \sigma_{\mu\nu} \gamma_5 \otimes \gamma^{\nu} \gamma_5
\end{array} \right )_{a b, c d}
= \left (
\begin{array}{llllllll}
{1\over4} & {1\over4} & {1\over4} & - {1\over4} & -{i\over4} &
{i\over4} & {i\over4} & {i\over4}
\\ {1\over4} & {1\over4} & -{1\over4} & {1\over4} & {i\over4} &
-{i\over4} & {i\over4} & {i\over4}
\\ {1\over4} & - {1\over4} & {1\over4} & {1\over4} & {i\over4} &
{i\over4} & -{i\over4} & {i\over4}
\\ -{1\over4} & {1\over4} & {1\over4} & {1\over4} & {i\over4} &
{i\over4} & {i\over4} & -{i\over4}
\\ {3i\over4} & -{3i\over4} & -{3i\over4} & -{3i\over4} & -{1\over4} &
-{1\over4} & -{1\over4} & {1\over4}
\\ -{3i\over4} & {3i\over4} & -{3i\over4} & -{3i\over4} & -{1\over4} &
-{1\over4} & {1\over4} & -{1\over4}
\\ -{3i\over4} & -{3i\over4} & {3i\over4} & -{3i\over4} & -{1\over4} &
{1\over4} & -{1\over4} & -{1\over4}
\\ -{3i\over4} & -{3i\over4} & -{3i\over4} & {3i\over4} & {1\over4} &
-{1\over4} & -{1\over4} & -{1\over4}
\end{array} \right )
\left (
\begin{array}{l}
\mathbf{1} \otimes \gamma^\mu
\\ \gamma^\mu \otimes \mathbf{1}
\\ \gamma_5 \otimes \gamma_\mu \gamma_5
\\ \gamma_\mu \gamma_5 \otimes \gamma_5
\\ \gamma^{\nu} \otimes \sigma_{\mu\nu}
\\ \sigma_{\mu\nu} \otimes \gamma^{\nu}
\\ \gamma^{\nu} \gamma_5 \otimes \sigma_{\mu\nu} \gamma_5
\\ \sigma_{\mu\nu} \gamma_5 \otimes \gamma^{\nu} \gamma_5
\end{array} \right )_{a d, b c}
\end{eqnarray}

\item
Products of two Dirac matrices with two anti-symmetric Lorentz
indices:
\begin{eqnarray}
\left (
\begin{array}{l}
\mathbf{1} \otimes \sigma_{\mu\nu} \gamma_5
\\ \gamma_5 \otimes \sigma_{\mu\nu}
\\ \sigma_{\mu\nu} \otimes \gamma_5
\\ \sigma_{\mu\nu} \gamma_5 \otimes \mathbf{1}
\\ \epsilon_{\mu\nu\rho\sigma} \sigma_{\rho l} \otimes \sigma_{\sigma l}
\\ \gamma_\mu \otimes \gamma_\nu \gamma_5 - (\mu \leftrightarrow \nu)
\\ \gamma_\mu \gamma_5 \otimes \gamma_\nu - (\mu \leftrightarrow \nu)
\\ \epsilon_{\mu\nu\rho\sigma} \gamma_\rho \otimes \gamma_\sigma
\\ \epsilon_{\mu\nu\rho\sigma} \gamma_\rho \gamma_5 \otimes \gamma_\sigma \gamma_5
\end{array} \right )_{a b, c d}
= \left (
\begin{array}{lllllllll}
{1\over4} & {1\over4} & {1\over4} & {1\over4} & {1\over4} &
{i\over4} & - {i\over4} & {1\over4} & - {1\over4}
\\ {1\over4} & {1\over4} & {1\over4} & {1\over4} & {1\over4} &
- {i\over4} & {i\over4} & - {1\over4} & {1\over4}
\\ {1\over4} & {1\over4} & {1\over4} & {1\over4} & - {1\over4} &
- {i\over4} & {i\over4} & {1\over4} & - {1\over4}
\\ {1\over4} & {1\over4} & {1\over4} & {1\over4} & - {1\over4} &
{i\over4} & - {i\over4} & - {1\over4} & {1\over4}
\\ 1 & 1 & - 1 & - 1 & 0 & 0 & 0 & 0 & 0
\\ - {i\over2} & {i\over2} & {i\over2} & - {i\over2} & 0 &
0 & 0 & {i\over2} & {i\over2}
\\ {i\over2} & - {i\over2} & - {i\over2} & {i\over2} & 0 &
0 & 0 & {i\over2} & {i\over2}
\\ {1\over2} & - {1\over2} & {1\over2} & - {1\over2} & 0 &
-{i\over2} & - {i\over2} & 0 & 0
\\ - {1\over2} & {1\over2} & - {1\over2} & {1\over2} & 0 &
-{i\over2} & - {i\over2} & 0 & 0
\end{array} \right )
\left (
\begin{array}{l}
\mathbf{1} \otimes \sigma_{\mu\nu} \gamma_5
\\ \gamma_5 \otimes \sigma_{\mu\nu}
\\ \sigma_{\mu\nu} \otimes \gamma_5
\\ \sigma_{\mu\nu} \gamma_5 \otimes \mathbf{1}
\\ \epsilon_{\mu\nu\rho\sigma} \sigma_{\rho l} \otimes \sigma_{\sigma l}
\\ \gamma_\mu \otimes \gamma_\nu \gamma_5 - (\mu \leftrightarrow \nu)
\\ \gamma_\mu \gamma_5 \otimes \gamma_\nu - (\mu \leftrightarrow \nu)
\\ \epsilon_{\mu\nu\rho\sigma} \gamma_\rho \otimes \gamma_\sigma
\\ \epsilon_{\mu\nu\rho\sigma} \gamma_\rho \gamma_5 \otimes \gamma_\sigma \gamma_5
\end{array} \right )_{a d, b c}
\end{eqnarray}

\end{enumerate}

%
\section{Chiral properties of Rarita-Schwinger fields}
\label{app:chiral_rep}
%

In this appendix, we study the chiral properties of Rarita-Schwinger
fields. As previously described in
Section~\ref{sec:chiral_representation}, we only need to study the
properties of $(LL)L$, $(LL)R$, $(LR)L$ and $(RL)L$.

%
\subsection{Chiral properties of $(LL)L$}
\label{app:LLL}
%

The chiral representations of $(LL)L$ are $(\mathbf{1},~\mathbf{1})
\oplus (\mathbf{8},~\mathbf{1}) \oplus (\mathbf{8},~\mathbf{1})
\oplus (\mathbf{10},~\mathbf{1})$. We will study them separately in
the following.

(1) In principle, there are eight possibilities of making the
Rarita-Schwinger fields as shown in Eq.~(\ref{eq:rariltordering}).
However, the chiral representation $(\mathbf{1},~\mathbf{1})$ has
just two non-zero fields:
\begin{eqnarray}
\begin{array}{l}
\Lambda_{L1\mu} = \epsilon_{abc} \epsilon^{ABC} (L_A^{aT} C L_B^b)
\gamma_\mu L_C^c \, ,
\\ \Lambda_{L2\mu} = \epsilon_{abc} \epsilon^{ABC} (L_A^{aT} C \gamma_5 L_B^b) \gamma_\mu \gamma_5
L_C^c \, .
\end{array}
\end{eqnarray}
Similarly performing the Fierz transformation to relate
$\Lambda_{Li\mu}$ and $\Lambda^{\prime}_{Li\mu}$, we obtain the
solution that all such kind of fields vanish.

(2) The chiral representation $(\mathbf{10},~\mathbf{1})$ has two
non-zero fields:
\begin{eqnarray}
\begin{array}{l}
\Delta^P_{L7\mu} = \epsilon_{abc} S_P^{ABC} (L_A^{aT} C
\sigma_{\mu\nu} L_B^b) \gamma^\nu L_C^c \, ,
\\ \Delta^P_{L8\mu} = \epsilon_{abc} S_P^{ABC} (L_A^{aT} C \sigma_{\mu\nu} \gamma_5 L_B^b) \gamma^\nu \gamma_5
L_C^c \, .
\end{array}
\end{eqnarray}
Similarly performing the Fierz transformation to relate
$\Delta^P_{Li\mu}$ and $\Delta^{P\prime}_{Li\mu}$, we obtain the
solution that all such kind of fields vanish.

(3) The chiral representation $(\mathbf{8},~\mathbf{1})$ has two
non-zero fields:
\begin{eqnarray}
\begin{array}{l}
N^{N}_{L1\mu} = \epsilon_{abc} \epsilon^{ABD} \lambda_{DC}^N
(L_A^{aT} C L_B^b) \gamma_\mu L_C^c \, ,
\\ N^{N}_{L2\mu} = \epsilon_{abc}
\epsilon^{ABD} \lambda_{DC}^N (L_A^{aT} C \gamma_5 L_B^b) \gamma_\mu
\gamma_5 L_C^c \, .
\end{array}
\end{eqnarray}
Similarly performing the Fierz transformation to relate
$N^{N}_{Li\mu}$ and $N^{N\prime}_{Li\mu}$, we obtain the solution
\begin{equation} \nonumber
N^{N}_{L2\mu} = N^{N}_{L1\mu} \, .
\end{equation}
Others are just zero. There is only one non-vanishing octet baryon
field.

%
\subsection{Chiral properties of $(LL)R$, $(LR)L$ and $(RL)L$}
\label{app:LLR}
%

The chiral representations of $(LL)R$, $(LR)L$ and $(RL)L$ are
$(\mathbf{\bar 3},~\mathbf{3}) \oplus (\mathbf{6},~\mathbf{3})$. We
will study them separately in the following.

(1) The chiral representation $(\mathbf{\bar 3},~\mathbf{3})
\rightarrow \mathbf{1_f}$ has two non-zero components:
\begin{eqnarray}
\begin{array}{l}
\Lambda_{M1\mu} = \epsilon_{abc} \epsilon^{ABC} (L_A^{aT} C L_B^b)
\gamma_\mu R_C^c \, ,
\\ \Lambda_{M2\mu} = \epsilon_{abc} \epsilon^{ABC} (L_A^{aT} C \gamma_5 L_B^b) \gamma_\mu \gamma_5
R_C^c \, .
\end{array}
\end{eqnarray}
Similarly performing the Fierz transformation to relate
$\Lambda_{Mi\mu}$ and $\Lambda^{\prime}_{Mi\mu}$, we obtain the
solution
\begin{equation} \nonumber
\Lambda_{M1\mu} = - \Lambda_{M2\mu} \, .
\end{equation}
Others are just zero. There is only one non-vanishing field. Others
$(LR)L$ and $(RL)L$ can be related to this one.

(2) The chiral representation $(\mathbf{6},~\mathbf{3}) \rightarrow
\mathbf{10_f}$ has two non-zero components:
\begin{eqnarray}
\begin{array}{l}
\\ \Delta^P_{M7\mu} = \epsilon_{abc} S^{ABC} (L_A^{aT} C \sigma_{\mu\nu} L_B^b) \gamma^\nu R_C^c \, ,
\\ \Delta^P_{M8\mu} = \epsilon_{abc} S^{ABC} (L_A^{aT} C \sigma_{\mu\nu} \gamma_5 L_B^b) \gamma^\nu \gamma_5
R_C^c \, .
\end{array}
\end{eqnarray}
Others are just zero. Similarly performing the Fierz transformation
to relate $\Delta^P_{Mi\mu}$ and $\Delta^{P\prime}_{Mi\mu}$, we
obtain the solution
\begin{equation} \nonumber
\Delta^P_{M7\mu} = - \Delta^P_{M8\mu} \, .
\end{equation}
There is only one non-vanishing field. Others $(LR)L$ and $(RL)L$
can be related to this one.

(3) The chiral representations $(\mathbf{\bar 3},~\mathbf{3})
\rightarrow \mathbf{8_f}$ has only two non-zero interpolators:
\begin{eqnarray}
\begin{array}{l}
N^{N}_{M1\mu} = \epsilon_{abc} \epsilon^{ABD} \lambda_{DC}^N
(L_A^{aT} C L_B^b) \gamma_\mu R_C^c \, ,
\\ N^{N}_{M2\mu} = \epsilon_{abc}
\epsilon^{ABD} \lambda_{DC}^N (L_A^{aT} C \gamma_5 L_B^b) \gamma_\mu
\gamma_5 R_C^c \, .
\end{array}
\end{eqnarray}
Similarly performing the Fierz transformation to relate
$N^{N}_{Mi\mu}$ and $N^{N\prime}_{Mi\mu}$, we obtain the solution
\begin{equation} \nonumber
N^{N}_{M1\mu} = -  N^{N}_{M2\mu} \, .
\end{equation}
In order to study the chiral representations
$(\mathbf{6},~\mathbf{3}) \rightarrow \mathbf{8_f}$, we need to
consider the second way (see the discussion in
Section~\ref{subsub:octet}) which leads to four non-zero fields:
\begin{eqnarray}
\begin{array}{l}
\widetilde{N}^{N}_{M1\mu} = \epsilon_{abc} \epsilon^{ACD}
\lambda_{DB}^N (L_A^{aT} C L_B^b) \gamma_\mu R_C^c \, ,
\\ \widetilde{N}^{N}_{M2\mu} = \epsilon_{abc}
\epsilon^{ACD} \lambda_{DB}^N (L_A^{aT} C \gamma_5 L_B^b) \gamma_\mu
\gamma_5 R_C^c \, ,
\\ \widetilde{N}^{N}_{M7\mu} = \epsilon_{abc}
\epsilon^{ACD} \lambda_{DB}^N (L_A^{aT} C \sigma_{\mu\nu} L_B^b)
\gamma^\nu R_C^c \, ,
\\ \widetilde{N}^{N}_{M8\mu} = \epsilon_{abc}
\epsilon^{ACD} \lambda_{DB}^N (L_A^{aT} C \sigma_{\mu\nu} \gamma_5
L_B^b) \gamma^\nu \gamma_5 R_C^c \, .
\end{array}
\end{eqnarray}
By using the Jacobi identity in Eq.~(\ref{eq:Jacobi}), we obtain:
\begin{equation}\nonumber
\widetilde{N}^{N}_{M1\mu} = {1 \over 2} N^{N}_{M1\mu}\,
,~\widetilde{N}^{N}_{M2\mu} = {1 \over 2} N^{N}_{M2\mu}\, .
\end{equation}
Similarly performing the Fierz transformation to relate
$\widetilde{N}^{N}_{Mi\mu}$ and $\widetilde{N}^{N\prime}_{Mi\mu}$,
we obtain the solution
\begin{eqnarray}\nonumber
\begin{array}{l}
\widetilde{N}^{N}_{M2\mu} = - \widetilde{N}^{N}_{M1\mu} = - {1 \over
2} N^{N}_{M1\mu}  \, , \\ \widetilde{N}^{N}_{M8\mu} = -
\widetilde{N}^{N}_{M7\mu} \, .
\end{array}
\end{eqnarray}
All together there are two non-vanishing independent fields:
$\widetilde{N}^{N}_{M1\mu}$ and $\widetilde{N}^{N}_{M7\mu}$.
$\widetilde{N}^{N}_{M1\mu}$ is related to $N^{N}_{M1\mu}$, and so
belongs to the chiral representation $((\mathbf{\bar
3},~\mathbf{3}))$. However, the other $\widetilde{N}^{N}_{M7\mu}$
can not be related to $N^N_{Mi\mu}$, and so contains some
$(\mathbf{6},~\mathbf{3})$ component. other Others $(LR)L$ and
$(RL)L$ can be related to $(LL)R$. Chiral properties of the tensor
fields can be also explored in completely the same procedure
explained here. Therefore, we do not show this case any more.

\end{document}